\documentclass[fleqn,usenatbib]{mnras}

\usepackage{newtxtext,newtxmath}

\usepackage[T1]{fontenc}
\usepackage{ae,aecompl}

\usepackage{graphicx}	
\usepackage{amsmath}	
\usepackage{xspace}
\DeclareGraphicsExtensions{.ps}

\newcommand{\planck}{\textit{Planck}}
\newcommand{\wmap}{\textit{WMAP}}

\newcommand{\camb}{{\small CAMB}}
\newcommand{\halofit}{{\small Halofit }}
\newcommand{\dm}{\textit{DM-only}}
\newcommand{\agn}{\textit{AGN}}

\newcommand{\Euclid}{\textsc{Euclid}}
\newcommand{\LSST}{\textsc{LSST}}
\newcommand{\NGRST}{\textsc{NGRST}}

\usepackage{subfiles}
\usepackage{aas_macros}
\usepackage{hyperref} 
\usepackage{multicol}
\hypersetup{colorlinks,citecolor=blue,linkcolor=blue,urlcolor=blue}

\title[Accuracy of the halo model
]{The BAHAMAS project: Evaluating the accuracy of the halo model in predicting the non-linear matter power spectrum}

\author[A. Acuto et al.]{
Alberto Acuto\thanks{E-mail: a.acuto@2017.ljmu.ac.uk},
Ian G. McCarthy\thanks{E-mail: i.g.mccarthy@ljmu.ac.uk}, 
Juliana Kwan, Jaime Salcido, Sam G. Stafford, Andreea S. Font\\
Liverpool John Moores University, 146 Brownlow Hill, Liverpool L3 5RF, UK\\
}

\date{Accepted XXX. Received YYY; in original form ZZZ}

\pubyear{2021}

\begin{document}
\label{firstpage}
\pagerange{\pageref{firstpage}--\pageref{lastpage}}
\maketitle

\begin{abstract}
The halo model formalism is widely adopted in cosmological studies for predicting the growth of large-scale structure in the Universe.  However, to date there have been relatively few direct comparisons of the halo model with more accurate (but much more computationally expensive) cosmological simulations. We test the accuracy of the halo model in reproducing the non-linear matter power spectrum, $P(k)$, when the main inputs of the halo model (specifically the matter density profiles, halo mass function, and linear bias) are taken directly from the BAHAMAS simulations and we assess how well the halo model reproduces $P(k)$ from the same simulations.  We show that the halo model generally reproduces $P(k)$ in the deep non-linear regime (1-halo) to typically a few percent accuracy, but struggles to reproduce (approx. 15\% error) $P(k)$ at intermediate scales of $0.1 \la k \ [h/{\rm Mpc}] \la 3$ at $z=0$, marking the transition between the 1-halo and 2-halo terms.  We show that the magnitude of this error is a strong function of the halo mass definition (through its effects on radial extent of haloes) and of redshift.  Furthermore, we test the accuracy of the halo model in recovering the \textit{relative} impact of baryons on $P(k)$.  We show that the systematic errors in recovering the absolute
$P(k)$ largely cancel when considering the relative impact of baryons.  This suggests that the halo model can make precise predictions for the baryonic suppression, offering a fast and accurate way to adjust collisionless matter power spectra for the presence of baryons and associated processes.
\end{abstract}

\begin{keywords}
gravitational lensing: weak -- dark matter -- large-scale structure of the Universe -- cosmology: theory
\end{keywords}


\section{Introduction} \label{introduction}

Large-scale structure (LSS) cosmology is now entering a golden era, with a large number of ongoing and forthcoming surveys poised to accurately measure the growth of structure over a wide range of physical scales.  For example, Stage-IV cosmic shear surveys such as \Euclid\footnote{\href{https://www.euclid-ec.org/}{https://www.euclid-ec.org/}},
the Rubin Observatory Legacy Survey of Space and Time (\LSST)\footnote{\href{https://www.lsst.org/}{https://www.lsst.org/}}, and the Nancy Grace Roman Space Telescope (\NGRST)\footnote{\href{https://roman.gsfc.nasa.gov/}{https://roman.gsfc.nasa.gov/}} aim to measure the matter power spectrum to percent level accuracy, in principle allowing constraints to be placed on important cosmological parameters, such as the dark energy equation of state, to a similar level of accuracy.  Comparatively tight constraints are also expected from forthcoming X-ray surveys with eROSITA\footnote{\href{https://www.mpe.mpg.de/eROSITA}{https://www.mpe.mpg.de/eROSITA}}, Sunyaev-Zel'dovich (SZ) effect surveys with SPT-3G\footnote{\href{https://pole.uchicago.edu/}{https://pole.uchicago.edu/}}, Advanced ACTPol\footnote{\href{https://act.princeton.edu/}{https://act.princeton.edu/}}, and Simons Observatory\footnote{\href{https://simonsobservatory.org/}{https://simonsobservatory.org/}}, and optical surveys (e.g., galaxy clustering, galaxy clusters) such as \LSST, \Euclid~and DESI\footnote{\href{https://www.desi.lbl.gov/}{https://www.desi.lbl.gov/}}.

In order to deliver on the aims of these surveys, a clear requirement is that we must be able to predict the observables (e.g., weak lensing power spectrum, galaxy clustering, SZ power spectrum) for a given cosmology, to an accuracy that is at least as precise as the statistical measurement errors. Otherwise, we risk biasing the derived cosmological parameters.  In the specific case of weak lensing, this means predicting the matter power spectrum to percent level accuracy.  The problem is particularly challenging, as most of the signal from current LSS tests comes from non-linear scales, thus requiring theoretical models to accurately follow matter as shells cross and collapse into `haloes', with galaxies potentially forming at their centres.

At present there are two general approaches to modelling the clustering and non-linear growth of matter: via direct N-body cosmological simulations or the so-called halo model.  In the former case, matter is discretised into large numbers of particles and their equations of motion are solved in the presence of a time-evolving background expansion.  Depending on the resolution and number of particles, such calculations can be computationally expensive and cannot at present be directly incorporated in, e.g., Markov chain approaches to cosmological parameter inference (which typically require thousands of evaluations).  A promising solution to this problem is via emulation techniques (e.g., \citealt{Kwan15,heitmann16,Rogers2019,DeRose2019,nishimichi19,pellejero20, spurio21, bose21}), whereby a grid of cosmological simulations spanning some cosmological parameter landscape is first run and then an emulator (e.g., based on Gaussian process interpolation or neural networks) is used to quickly and accurately interpolate the results (e.g., the matter power spectrum) for any choice of cosmological parameters that are within the boundaries of the initial suite.  Such emulators, which can typically be run in fractions of a second, can be implemented in cosmological likelihood analyses.   

While emulation of cosmological simulations is clearly going to be an important tool going forward, it does have limitations.  For example, predictions are confined to the parameter space defined in the initial base grid of simulations, where there is a trade off between accuracy of the emulator prediction, the volume of the cosmological landscape being surveyed (i.e., the range of parameter values included), and the number of simulations that can feasibly be run from the base grid.  In addition, adding new extensions (e.g., beyond $\Lambda$CDM) or probing a larger (or different) cosmological parameter spaces often requires one to considerably adapt the base grid of simulations used to build the emulator, which can be computationally expensive.  In addition, at the moment most emulators are based on simulations that do not incorporate the important role of baryons.  Nevertheless, it has been shown in recent work based on cosmological hydrodynamical simulations that baryons can alter the matter power spectrum by up to a few tens of percent (e.g., \citealt{jing2006, vandaalen11, schneider2015, Mummery17, chisari19, vanDaalen2020}), which is significantly larger than the anticipated statistical error of future weak lensing measurements.

The halo model \citep{peacock2000,Seljak00,ma2000,Cooray02, smith03} potentially provides a solution to many of these issues.  In brief, the halo model provides a simple, physically-motivated picture for the clustering of matter and haloes.  In its standard and simplest form, the halo model requires as input the distribution of matter within haloes (i.e., their density profiles), the mass function of haloes (i.e., the abundance of haloes as a function of mass and redshift), and a prescription for halo bias which describes how the clustering of haloes is related to the clustering of matter in general.  The mass function and bias can in principle be derived from analytic/semi-analytic arguments (e.g. \citealt{PS1974,Sheth99,Sheth2001}), though it is now commonplace to use large cosmological simulations to provide more accurate determinations of these quantities \citep{Tinker08, despali16, bocquet16, castro2020}.  The density profiles are generally also extracted from cosmological simulations (e.g., a Navarro-Frenk-White profile using a mass--concentration--redshift relation).  

As the halo model can be written down analytically in a small number of equations (see Section \ref{methods}), it can be evaluated extremely quickly and therefore easily incorporated within cosmological pipelines.  Furthermore, as it provides a physically-intuitive description for the matter distribution within haloes, it is relatively straightforward to adjust it to incorporate the impact of baryons (e.g. \citealt{semboloni2011, semboloni2013, Fedeli14,debackere2019, mead2020}).  The free parameters associated with the baryon physics can either be constrained by cosmological hydrodynamical simulations, external observations, or marginalised over when jointly fitting a cosmological dataset (e.g., cosmic shear) and baryonic parameters (e.g. \citealt{Shirasaki2019}).

Given its speed, flexibility, and intuitive design, the halo model is used for theoretical interpretation in many cosmological surveys (e.g., \citealt{Batt2012_SZ,Horowitz,Hill14, robertsonN20,schneider19, giocoli2020}).  However, an important question is how accurate is the halo model?  In particular, in order to derive unbiased constraints on cosmological (and possibly baryonic) parameters, we require that the halo model predicts the non-linear matter power spectrum ($P(k)$) accurately given the input profiles, mass function, and linear bias.  However, to our knowledge, there have been very few (and no recent, in the era of precision cosmology) tests of the internal accuracy of the halo model for predicting $P(k)$.  By `internal' accuracy, we mean the following: given the density profiles, mass function, and linear bias from a particular simulation, how well does the halo model reproduce the measured power spectrum from the same simulation?  A second important question is, how do baryons change this picture?  Although these questions are relatively simple, they are in fact very challenging tests of the halo model, as once the input profiles, mass function, and bias are specified, there are no free parameters in the standard halo model.  It is possible to add extra degrees of freedom to the halo model and to constrain these using fits to the power spectra of cosmological simulations (e.g, \citealt{Mead15,Mead16,mead2020hx}), but the physical interpretation of such additions is unclear, as are the potential dependencies of these terms on the cosmological parameters.

Given that the halo model can be informed using quantities extracted directly from the simulations, should we not expect it to accurately recover the matter clustering in the simulations?  In terms of the density profiles, one source of error is that scatter in the density profiles at fixed halo mass and redshift is generally not incorporated into the halo model.  Additionally, the model ignores issues such as non-sphericity of haloes (e.g. \citealt{smith2005}) and the presence of substructures (and their clustering, which may differ from the smooth dark matter component, e.g. \citealt{sheth2003}).  There is also an expectation that the assumption of linear bias will break down on certain scales (e.g., \citealt{smith2007,baldauf2012, meadverde}).  In addition, although the halo mass function is a simple statistic with apparently little wiggle room, we will show that the accuracy of the halo model actually depends strongly on the choice of halo mass definition.  

In the present study, we use the BAHAMAS\footnote{\href{https://www.astro.ljmu.ac.uk/~igm/BAHAMAS/}{https://www.astro.ljmu.ac.uk/~igm/BAHAMAS/}} simulations \citep{BAHAMAS, IGM18} to test the internal accuracy of the standard halo model in terms of its prediction for $P(k)$.  We evaluate the accuracy both in the context of collisionless physics (`dark matter only') and in the presence of baryons and processes associated with galaxy formation (e.g., feedback).  Lastly, we comment on the relative accuracy of the halo model, in terms of the ratio of the matter power spectrum in a hydrodynamical context to that from a collisionless context (sometimes referred to as the baryon `suppression factor').  

In this paper we adopt a \wmap~9 cosmology \citep{hinshaw13} with parameters $h=0.7$, $\Omega_{m}=0.2793$, $\Omega_b=0.0463$, $n_s =0.972$, $\sigma_{8}=0.8211$ and $\Omega_{\nu}=0.0$ .

The paper is structured as follows.  In Section \ref{methods} a general description of the halo model formalism and the BAHAMAS suite of simulations is provided. In Section \ref{calibration} we calibrate the halo model using the simulations and present tests of our methodology. In Section \ref{pk_exploration} we present the $P(k)$ predictions for both the collisionless and hydrodynamical cases, commenting on the absolute and relative accuracy of the halo model and discussing the implications of the results. Finally, in Section \ref{conclusions} we summarise our findings. 

\section{Methodology} \label{methods}

In this section we present a brief description of the standard halo model and how it is used to predict the 3D matter power spectrum, $P(k)$.  We refer the reader to the original studies that introduced this formalism \citep{peacock2000,Seljak00,ma2000,Cooray02, smith03} for further details (see also \citealt{mead2020} for an excellent recent discussion).

The halo model describes the clustering of haloes and matter via the power spectrum (i.e., the Fourier transform of their two-point correlation functions) as the sum of two terms, the so-called `1-halo' and `2-halo' terms:
\begin{equation}
P(k)^{\rm tot} = P(k)^{\rm 1h} + P(k)^{\rm 2h} \ \ \, 
\label{cl}
\end{equation}
\noindent where the first term (1-halo) describes the clustering of matter within a single halo (also called intra-halo clustering), while the second term (2-halo) describes the (correlated) clustering of matter in neighbouring haloes \citep{smith11}.

The 1-halo and 2-halo terms can be recast in terms of physical quantities as:
\begin{equation}
    \begin{gathered}
    P(k)^{\rm 1h} =  \int dM \frac{dn(M,z)}{dM} |\tilde{\mathit{X}}_{k}(M,z)|^{2} \ \ \ , \\
 P(k)^{\rm 2h} =  P_{\rm lin}(k,z) \left[ \int dM\frac{dn(M,z)}{dM} b(M,z) |\tilde{\mathit{X}}_{k}(M,z)|\right]^{2} \ \ \ , \\
\end{gathered}
\label{matterpowerspec}
\end{equation}
\noindent where $dn/dm$ is the halo mass function (the space density of haloes of a given mass), $b$ is linear halo bias computed via the square of the ratio of the halo power spectrum over the matter power spectrum in the linear regime, $P_{\rm lin}$ is the linear matter power spectrum, computed here using the software \camb~ \citep{LewisChallinor} \footnote{\href{https://camb.info/}{https://camb.info/}}, and $\tilde{\mathit{X}}_{k}$ is the Fourier-transform of the 3D spherical matter density profile convolved with the Fourier-transform of the top-hat window function, expressed as:
\begin{equation}
    \tilde{\mathit{X}}_{k}(k,M) = \frac{1}{\overline{\rho}}\int_{0}^{R_{\Delta}}  4\pi r^2 \rho(r, M, z)\frac{\sin(kr)}{kr}dr \ \ \ .
    \label{eq:windowfunc}
\end{equation}{}

Above, $\rho$ is the total matter density profile, $R_{\Delta}$ is the radial extent of the halo which is specified by the choice of halo mass definition (see discussion in Section \ref{profiles}), and $\overline{\rho}$ is the mean comoving density of the Universe. 

Below we will explore in detail each component of the halo model.

\subsection{Halo density profiles}

A key component in the halo model's prediction for $P(k)$ is the way that matter is distributed inside haloes; i.e., their total matter density profiles. A common choice in this regard, which is motivated on the basis of collisionless (N-body) cosmological simulations, is the Navarro-Frenk-White (NFW) profile \citep{NFW}:

\begin{equation}
\rho = \frac{\rho_0}{\biggl(\frac{r}{r_s}\biggr)\biggl(1+\frac{r}{r_s}\biggr)^2} \ \ \ ,
\label{eq:nfw}    
\end{equation}
\noindent where $r_s$ is the scale radius and $\rho_0$ is the normalisation.  The scale radius is a free parameter, whereas one can either leave the normalisation ($\rho_0$) free or specify it through the halo mass definition (e.g., chosen so that the mean density within $R_{200,{\rm crit}}$ from the simulations is 200 times the critical density).  The scale radius is often recast in terms of the halo concentration, $c_\Delta \equiv R_\Delta/r_s$, where $R_\Delta$ is the radius used in the halo mass definition.  The concentration is known to depend on halo mass, redshift, and the choice of cosmological parameters and various fitting functions for this behaviour have been proposed \citep{duffy08,diemer15, ludlow14, ludlow16}.  Using these fitting functions for the concentration, one completely specifies the distribution of mass within haloes given a total halo mass, redshift, and the cosmological parameters.

While the NFW profile provides a reasonably good description of the typical density profiles of collisionless simulations, it performs less well in describing the total matter density profiles in cosmological hydrodynamical simulations \citep{Duffy2010,dutton14,sereno16, schaller15a, schaller15b}.   One can generalise the NFW form to allow for additional freedom \citep{Nagai2007}: 
\begin{equation}
\rho(r,M,z) = \rho_{0}\left(\frac{r}{r_s}\right)^{\alpha}\left[1+\left(\frac{r}{r_s}\right)^{\gamma}\right]^{-\beta} \ \ \ ,    
\end{equation} 
\noindent where $\rho_{0}$, $\alpha$,$\gamma$ and $\beta$ are free parameters.  This parametric form is often used to model the pressure distribution of the hot gas around groups and clusters (e.g., \citealt{arnaud10,Batt2012_SZ}) but would also be suitable for the mass density distribution.  In principle the free parameters of the generalised NFW form are also functions of mass and redshift, which leads to an even larger number of free parameters which would be expected to have significant degeneracies. 

Our approach is to allow for additional freedom relative to the original NFW form, but with fewer free parameters than in the generalised NFW case.  In particular, we adopt the so-called Einasto profile \citep{Einasto}, which recent work has shown better reproduces the matter distribution in haloes in collisionless simulations \citep{springel2008,Navarro10,dutton14, brown2020}.  This is due to its additional flexibility relative to NFW (it has an additional free parameter) which ought to allow it to better describe hydrodynamical simulations as well (indeed we show this below, in Section \ref{pk_exploration}).  The Einasto profile can be expressed as:
\begin{equation}   
    \rho(r, M,z) = f_{0}(M,z)\exp{\left[-A(M,z) r^{\alpha(M,z)}\right]} \ \ \ ,
    \label{Einasto_eq}
\end{equation}
with three main parameters $f_{0}$, $A$ and $\alpha$ which need to be fit for.  As discussed in Section \ref{profiles}, for these three parameters we adopt power law dependencies on halo mass and redshift, resulting in a total of 9 free parameters overall to describe $\rho(M,r,z)$.

Note that, as our aim is primarily to test the accuracy of the halo model, one does not actually need to use a parametric form for the density profiles, but can instead use non-parametric (tabulated) density profiles directly from the simulations.  Indeed, we will show the results for both cases: fits to the profiles (with an Einasto form) and using the tabulated profiles directly.

\subsection{Halo mass function} \label{hmf_theory}

Another key ingredient of the halo model formalism is the halo mass function (HMF).  This quantity can be derived from analytic/semi-analytic theoretical arguments, such as those put forward by \citet{PS1974} and \citet{Sheth2001}.  However, more accurate representations can be derived from fits to large suites of cosmological simulations (e.g., \citealt{jenkins01,Tinker08,bocquet16,despali16,bocquet20,diemer20}).

It is commonplace to parametrise the halo mass function from cosmological N-body simulations as:
\begin{equation}
    \frac{dn(M,z)}{dM} = f(\sigma)\frac{\overline{\rho}}{M}\frac{\ln\sigma^{-1}}{dM},
    \label{eq:hmf}
\end{equation}
\noindent where $f(\sigma)$ is a function fit to the simulations to encapsulate the dependence on redshift and $\sigma$, the mass variance, is defined as:
\begin{equation}
    \sigma^2=\frac{1}{2\pi^2}\int P(k)\hat{W}^2(kR)k^2dk,
    \label{sigma_hmf}
\end{equation}
where $P(k)$ is the linear matter power spectrum, $\hat{W}$ is the Fourier transform of the top-hat window function.  This form has been shown to reproduce the halo mass function from simulations\footnote{The accuracy of the \citet{PS1974} and \citet{Sheth2001} mass functions is typically 20\%, with a general over-prediction of the abundance of the most massive objects \citep{Mead15, delpopolo17}.} to $\approx$10\% accuracy \citep{Tinker08,diemer20}.  Note that the cosmology dependence of the HMF enters in through both the mass variance (which depends on the cosmology-dependent linear power spectrum) and the mean density.

Note that these halo mass functions are typically derived from collisionless (\dm) simulations and therefore they do not account for any baryonic processes (e.g., feedback) affecting haloes.  Given the important role of baryons in setting the mass distributions of haloes, we will make use of the BAHAMAS HMFs to build a `correction' function to allow us to study the impact of baryons on the matter power spectrum via the halo model (see Section \ref{hmf_baryon_correction}).

As in the case of the density profiles, it is not necessary to use `off the shelf' predictions for the halo mass function for the present study.  We will therefore compare the predictions of the halo model using the \citet{Tinker08} (which is perhaps the most common choice in the literature) with those using the tabulated halo mass functions derived directly from the BAHAMAS simulations.  In this way we can assess the impact of using an internally consistent evaluation of the halo model.

\subsection{Halo bias and linear power spectrum} \label{bias_theory}

With the matter distribution within haloes (density profiles) and the number density (mass function) specified, the remaining ingredient is to specify how haloes cluster in space, in terms of their 2-point correlation function (or, in Fourier space, their power spectrum).  We adopt the standard assumption that haloes are linearly-biased tracers of the overall matter distribution.  Specifically, the linear bias is evaluated as:
\begin{equation}
    b(k) = \sqrt{\frac{P_{\rm hh}(k)}{P_{\rm lin,mm}(k)}},
\end{equation}
\noindent where $P_{\rm hh}(k)$ and $P_{\rm lin, mm}(k)$ are the halo-halo and linear matter-matter power spectra, respectively.  On large (linear) scales (small $k$), $b(k) \rightarrow$ const., which is what is typically referred to as just the linear bias, which is what we use here.

In the present study, we use the linear bias--peak height relation of \citet{Tinker10} to calculate the linear bias as a function of halo mass and redshift (and cosmological parameters).  Note that the peak height is defined as $\nu = \delta_{\rm crit}/\sigma(M)$ where $\delta_{\rm crit}$ is the density threshold for collapse (usually assumed to be equal to $1.686$) and $\sigma(M)$ is the linear matter variance measured within the Lagrangian scale, R, corresponding to halo mass $M$. 

Note that we could also use the linear bias from the BAHAMAS simulations, instead of that from \citet{Tinker10}.  However, the choice of bias does not turn out to be important for computing the total \textit{matter} power spectrum, $P(k)$, since, as we discuss below, an additive correction to the 2-halo term must be applied anyway (for unresolved haloes) in order to force it to match the linear theory prediction for the matter power spectrum on large scales.  This implies that, even if the adopted bias model were to be relatively inaccurate in terms of describing the simulations, the additive correction factor would compensate\footnote{A subtle point is that the correction factor is an additive term to the overall 2-halo term, whereas the bias enters into the 2-halo term in a multiplicative way and it depends on halo mass.  Thus, the additive correction term is not perfectly degenerate with the bias, but it is very close to being so.} by forcing the computed 2-halo term to match the linear prediction on large scales anyway.  If we were interested in examining the clustering of haloes (rather than matter), the choice of bias would obviously be much more important.  Furthermore, we note that the impact of baryons on the linear bias is expected to be small (e.g., \citealt{castro2020}), which we have verified with BAHAMAS, and therefore we use the same formalism \citep{Tinker10} for both the collisionless and baryon cases. 

A physical condition that must be met is that, when integrated over all halo masses, the bias must be unity.  That is, the total matter is unbiased, by definition; i.e.:

\begin{equation}
\int b(\nu)f(\nu)d\nu = 1 \ \ \ ,
\label{bias_nu}
\end{equation}
where $f(\nu) d\nu= (dn(M)/dM) (M / \overline{\rho}_m) dM$ \footnote{The conservation of matter requires fulfilling $\int b(M)M(dn(M)/dM) dM=\overline{\rho}$, where $\overline{\rho}$ is the mean background density.}. This is true when integrating over all possible halo masses.  However, due to finite box size and resolution, simulations do not sample all possible halo masses (in particular masses below resolution limit of the simulations), which means that if one integrates over just the resolved haloes in a simulation box, one does not recover the large-scale linear power spectrum using the halo model \citep{vanDaalen2015, schmidt16,mead2020hx}.  A correction for these unresolved haloes is therefore required to force the halo model to agree with linear theory on large scales.

We follow previous studies \citep{schmidt16,mead2020hx,philcox2020} and add in the contribution of low-mass haloes to the 2-halo term in order to recover the linear regime (see the appendix of \citealt{mead2020hx} for further discussion). 
From eqn.~\ref{bias_nu}, the additive term can be derived as:
\begin{equation}
    A_{\rm low} = 1-\frac{1}{\overline{\rho}}\int_{M_{\rm min}}^{\infty} b(M) \frac{dn(M)}{dM} dM \ \ \ ,
    \label{alow}
\end{equation}
where $M_{\rm min}$ is the minimum halo mass resolved in the simulation, which for BAHAMAS we take to be $4\times10^{11}\,$M$_{\odot}\,h^{-1}$.  

With the above we construct the additive component for the 2-halo term as:
\begin{equation}
    C_{\rm add} = \frac{A_{\rm low}\tilde{X}_{k}(M_{\rm min})}{M_{\rm min}} \ \ \ ,
    \label{cadd}
\end{equation}
where $\tilde{X}_{k}$ is the Fourier transform of the 3D density profile of the lowest resolved halo mass, $M_{\rm min}$. 

The term in eqn.~\ref{cadd} is added to the standard 2-halo term before being multiplied by the linear matter power spectrum, as:
\begin{equation}
 \begin{split}
 P(k)^{\rm 2h} =  \left[\int_{M_{\rm min}}^{\infty} \frac{dn(M,z)}{dM} b(M,z) |\tilde{\mathit{X}}_{k}(M,z)|dM+C_{\rm add}(M_{\rm min})\right]^{2}\\
 \times P_{\rm lin}(k,z).
\end{split}
\label{small_haloes}
\end{equation}

As noted above, this approach guarantees that the constructed halo model reproduces the linear clustering of matter on large scales.  Alternatively, one could simply replace the 2-halo term with the linear power spectrum and find very similar results.  The only significant difference between eqn.~\ref{small_haloes} and the linear power spectrum occurs on small scales where the 1-halo term is already dominant.  But for completeness we use the full expression for the (re-normalised) 2-halo term.

\subsection{Additional (ad-hoc) considerations}

In this section we will briefly explore some additional, ad-hoc adjustments of the standard halo model which have been implemented in previous works.  Specifically, we follow some of the adjustments introduced in \citet{Mead15} to avoid several unphysical artefacts in the standard halo model.

Firstly, we apply a smooth cut-off of the 2-halo term on quasi-linear scales.  As discussed by \citet{Mead15} and \citet{Mead16}, linear theory overpredicts the matter power spectrum on quasi-linear scales and does not accurately capture the damping of the baryonic acoustic oscillations (BAO) peaks in particular $k\approx 0.2-0.4\,h/$Mpc.  Following \citet{Mead15} (see their section 3.2.1) we therefore apply a tapering to the 2-halo term on quasi-linear scales using:
\begin{equation}
\Delta^{'2}_{\rm 2h}(k) = \left[1-f \tanh^{2}(k\sigma_{\rm v, d}/\sqrt{f})\right]\Delta_{\rm 2h}^{2}(k) \ \ \ ,
\label{1h_2h_trans}
\end{equation}
where $\Delta^2$ is the dimensionless power spectrum computed via $\Delta^2(k) = 4\pi (k/2\pi)^3 P(k)$ and $\sigma_{v}$ is the $1$D linear-theory displacement variance defined as:
\begin{equation}
\sigma_{\rm v,d}^2 = \frac{1}{3}\int_{0}^{\infty} \frac{\Delta^2_{\rm lin}(k)}{k^3}dk \ \ \ ,
\end{equation}
where $f$ in eqn.~\ref{1h_2h_trans} is the damping factor, and $\Delta_{\rm lin}^{2}$ is the dimensionless linear power spectrum computed using \camb~. \citet{Mead16} find that $f$ has a small dependence on $\sigma_{\rm v, d}$ as $f= 0.095\sigma_{\rm v,d}^{1.37}$.  Note that the application of eqn.~\ref{1h_2h_trans} only affects $P(k)$ by about a percent on large scales and therefore has no significant impact on the results or conclusions of our study, but we include it for completeness.

As also discussed by \citet{Mead15}, but firstly presented in \citet{cooray2002} (see also \citealt{smith11, valageas2011}), the standard (unmodified) 1-halo term displays unphysical behaviour at very large scales.  In short, the 1-halo term becomes larger than that predicted by linear theory on very large scales, because the halo model implicitly assumes that haloes are randomly distributed on large scales when, in reality they are clustered and distributed more smoothly than random.  Following \citet{Mead15} (see their section 3.2.2), we truncate the 1-halo term on large scales using:
\begin{equation}
    \Delta_{\rm 1h}^{'2}= \left[1- e^{-(k/k_{*})^2}\right]\Delta_{\rm 1h}^{2} \ \ \ .
    \label{1h_kstar}
\end{equation}

This ad-hoc correction suppresses 1-halo power at scales $k \la k_*$.  Mead et al.~find the value of $k_{*}$ depends on the $1$D linear-theory displacement variance as $k_{*} = 0.548 \sigma_{\rm v, d}^{-1}(z)$. 

In addition to the above modifications, \citet{Mead15} (see also \citealt{mead2020hx,mead2020}) consider a number of other modifications of the halo model designed to provide a better fit to the non-linear matter power spectra of cosmological simulations.  While allowing for extra degrees of freedom does allow the halo model to provide an improved fit to the simulations, one could argue that in doing so we are sacrificing the physical intuitiveness of the model for new parameters whose interpretation is ambiguous.  Whether these parameters should depend on baryon physics or cosmology is also unclear.  Therefore, as discussed in the Introduction, we take a different approach and simply evaluate the accuracy of the standard (unmodified, modulo that mentioned above) halo model and assess to what extent it can be reliably applied in this era of precision large-scale structure cosmology.

\subsection{BAHAMAS simulations}
\label{simulations}

To test the accuracy of the halo model, we employ the BAHAMAS suite of cosmological hydrodynamical simulations \citep{BAHAMAS, IGM18}, as well as their collisionless (\dm) counterparts.  Most of the BAHAMAS runs consist of $400$ Mpc$h^{-1}$ comoving on a side, periodic boxes containing $2\times1024^3$ particles.  For the fiducial WMAP9 run from \citet{BAHAMAS} that we use, the dark matter particle mass is $3.85\times10^{9}\,$M$_{\odot}\,h^{-1}$ and the initial gas mass is $7.66\times10^{8}\,$ M$_{\odot}\,h^{-1}$. 

The Boltzmann code \camb~(\citet{camb}, version April 2014) was used to compute the transfer functions which were supplied to a modified version of the N-GenIC\footnote{\href{https://github.com/sbird/S-GenIC}{https://github.com/sbird/S-GenIC}} code to create the initial conditions, at a starting redshift of $z=127$.  The N-GenIC code was modified to include second-order Lagrangian Perturbation Theory (2LPT) and support for massive neutrinos, although note that most of our tests use the fiducial BAHAMAS simulation from \citet{BAHAMAS} which has massless neutrinos and a WMAP9 maximum-likelihood cosmology \citep{hinshaw13}.

The BAHAMAS simulations were run with the Lagrangian TreePM-SPH code {\small GADGET-3} \citep{gadget3}.  Subgrid physics developed for the OWLS project (Overwhelmingly Large Simulations \citealt{Schaye10, lebrun14}) was included in the hydrodynamical simulations, specifically prescriptions for metal-dependent radiative cooling, star formation, stellar evolution, chemical enrichment, stellar feedback, and black hole growth and AGN feedback (see \citealt{Schaye10} and references therein).  While in OWLS (and cosmo-OWLS) no attempt was made to calibrate the feedback parameters to reproduce observations, the approach of BAHAMAS was to explicitly calibrate the efficiencies of the stellar and AGN feedback to reproduce the local ($z\approx0$) galaxy stellar mass function and the gas fractions of galaxy groups and clusters.  The objective in doing so was to ensure that the most massive haloes (massive galaxies up to clusters), which contribute the most to the matter power spectrum \citep{vanDaalen2015,mead2020hx}, have the correct baryon fractions.  \citet{vanDaalen2020} have shown that the baryon fraction on the group scale ($M \sim 10^{14}$ M$_\odot\,h^{-1}$) is the key quantity in determining the impact of baryon physics on the matter power spectrum.

A standard Friend-Of-Friends (hereafter {\small FOF}) algorithm with a linking length of $b=0.2$ times the mean interparticle separation is run to identify FOF haloes, from which we calculate the halo mass function and the mean matter density profiles (see Section \ref{calibration}).  Note that because we first identify haloes with a {\small FOF} algorithm and then compute their SO masses, haloes cannot overlap spatially.  For each FOF halo, we extract all of the particles in a sphere of radius $5 R_{200,{\rm crit}}$ centred on the most bound particle for calculation of the density profiles.  In practice, the outer radius we use for the density profiles depends on the adopted halo mass definition, but our choice for storing the particles around haloes is conservatively large.

While we use a fiducial WMAP9 cosmology to test the halo model, we note that the BAHAMAS simulations also contain extensions to the standard model, including massive neutrinos \citep{Mummery17}, a running of the scalar spectral index \citep{stafford2019,stafford2020} and dynamical dark energy \citep{pfeifer2020}.  In future work, we will explore whether our findings here also extend to these `non-standard' cosmologies. 

We will refer to the full hydrodynamics case as \agn~and the collisionless case as \dm, and we consider the total matter (dark matter+baryons) power spectrum.

\section{Informing the halo model with BAHAMAS}
\label{calibration}

In this section we extract the `ingredients' necessary to evaluate the halo model (namely the matter density profiles and the halo mass function) from the BAHAMAS simulations.

\subsection{Matter density profiles}
\label{profiles}

Here we describe our procedure for creating stacked total mass density profiles from BAHAMAS for use in the halo model.  We select {\small FOF} haloes of mass $10^{11}< M_{\Delta}[$M$_{\odot}\,h^{-1}]\approx 5\times 10^{15}$ and from $0 \leq z \leq 3$, where the halo mass, $M_\Delta$, is defined according to one of the four mass definitions that we present below (see Table \ref{tab:fittings_prof}).  We adopt a (logarithmic) halo mass bin width of $d\log_{10}M=0.125$.  We extract all of the particles belonging to all of the haloes in a given mass bin, depositing them into $\approx150$ spherical shells (scaled by $R_\Delta$ and centred by the halo centre of potential), logarithmically-spaced from $10^{-3}R_{\Delta}$ to $R_{\Delta}$, where $\Delta$ is generalised to cover the four different mass definitions that we explore (see Table \ref{tab:fittings_prof}).  For a given radial bin, we compute the mass-weighted mean radius: 
\begin{equation}
    r_{w} = \frac{\Sigma_{i}m_{i}r_{i}}{\Sigma_{i}m_{i}}.
    \label{weighted_radius}
\end{equation}

Note that we are able to reach such small inner radii ($10^{-3}R_{\Delta}$) because we are considering all the particles in many haloes stacked together.  Note also that for the lowest mass haloes we consider that $10^{-3}R_{\Delta}$ can actually probe scales below the softening length of the simulations ($4$ pkpc$\,h^{-1}$), but this does not effect our ability to evaluate the consistency of the halo model, since the softening will also effect the power spectrum, $P(k)$, in the same way.

As already noted, we will use both the tabulated density profiles directly and parametric fits to those profiles, using an Einasto form.  To allow for potential halo mass and redshift dependencies of the three main parameters in the Einasto profile ($f_{0},A$ and $\alpha$, see eqn.~\ref{Einasto_eq}), we model them with a simple power law dependence on both quantities, for example:
\begin{equation}
    f_{0}(M,z) = f_{0,\rm int} \left(\frac{M}{M_{\rm ref}} \right) ^{f_{\rm m}}(1+z)^{f_{z}}, \ \ \
    \label{eq:parms}
\end{equation}
\noindent where $M_{\rm ref}$ is a reference mass (or pivot point) used for normalisation of the function, which we adopt as $10^{13}$ M$_\odot\,h^{-1}$. 

We determine the best-fit parameters using a nonlinear least-squares Levenberg-Marquardt approach \citep{Markwardt09} with the {\small IDL} routines {\small CURVEFIT} and {\small MPCURVEFIT} using the partial derivatives with respect to each parameter to help the convergence of the fit.  We simultaneously fit to the stacked density profiles over the full range of radial bins, halo mass bins, and redshifts described above.  Note that since a given radial bin typically contains large numbers of particles, the Poisson uncertainties are typically negligibly small.  Therefore, we simply neglect these uncertainties, giving equal weight to each radial bin in the fit.

In Table \ref{tab:fittings_prof} we present the best-fit parameters for the \dm~and \agn~total matter density profiles for the four different mass definitions.  We note that there are likely to be large degeneracies between the derived parameters, but this is generally unimportant for our purposes, since we only require that the function provides a good fit to the simulated profiles for the range of halo masses, radii, and redshifts that we consider.  Because of the degeneracies between the parameters, the best-fit values themselves do not necessarily have important physical significance.

\begin{table*}
\begin{tabular}{lcc|ccc|ccc|ccc}
\hline
Overdensity & Type&$M_{\rm ref} [$M$_{\odot}\,h^{-1}]$ &$f_{0, \rm int} [\times10^{9}]$ &$f_{m}$&$f_{z}$&$A_{0}$&$a_{m}$&$a_{z}$&$\alpha_{0}$&$\alpha_{m}$&$\alpha_{z}$ \\
\hline
$\Delta =200m$ &\dm~&$10^{13}$ & $0.3474$ & $-0.0056$ & $-1.804$ & $14.01$ & $-0.00475$ & $-0.1891$ & $0.298$ & $-0.0182$ & $-0.053$ \\
$\Delta =200c$ &\dm~&$10^{13}$ & $0.2295$ & $0.2633$ & $-1.196$ & $12.04$ & $0.01594$ & $-0.0062$ & $0.255$ & $-0.0522$ & $-0.161$ \\
$\Delta =500m$ &\dm~&$10^{13}$ & $0.3395$ & $0.0025$ & $-1.624$ & $12.77$ & $-0.00371$ & $-0.1861$ & $0.242$ & $-0.0218$ & $-0.096$ \\
$\Delta =500c$ &\dm~&$10^{13}$ & $0.1354$ & $0.3520$ & $-0.869$ & $10.42$ & $0.0259$ & $0.0213$ & $0.279$ & $-0.0711$ & $-0.239$ \\

\hline
$\Delta =200m$ &\agn~&$10^{13}$ & $0.5083$ & $-0.099$ & $0.642$ & $16.63$ & $-0.0115$ & $0.006$ & $0.189$ & $-0.019$ & $-0.025$ \\
$\Delta =200c$ &\agn~&$10^{13}$ & $14504.6$ & $-0.263$ & $-1.76$ & $25.29$ & $-0.0147$ & $-0.023$ & $0.105$ & $-0.032$ & $-0.104$ \\
$\Delta =500m$ &\agn~&$10^{13}$ & $12491.88$ & $-0.316$ & $-2.342$ & $25.52$ & $-0.0173$ & $-0.118$ & $0.104$ & $-0.027$ & $-0.115$ \\
$\Delta =500c$ &\agn~&$10^{13}$ & $35538.27$ & $-0.287$ & $-1.793$ & $25.09$ & $-0.0152$ & $-0.025$ & $0.096$ & $-0.033$ & $-0.126$ \\
\hline
\end{tabular}
\caption{Best-fit Einasto parameter values (see eqns.~\ref{Einasto_eq} and \ref{eq:parms}) describing the density profiles of the BAHAMAS \dm~and \agn~cases for four different halo mass definitions.}
\label{tab:fittings_prof}
\end{table*}

In Table \ref{tab:fittings_prof} we have introduced the four different halo mass definitions that we consider, corresponding to spherical overdensities of either 200 or 500 times either the critical or mean density of the universe at a given redshift, where the critical density, $\rho_{\rm crit}(z)$, is defined as $3H(z)^2/8\pi G$ and the mean density is just $\Omega_m(z) \rho_{\rm crit}(z)$. We fit to the mass density profiles normalised by either $\rho_{\rm mean}$ or $\rho_{\rm crit}$ (depending on the halo mass definition), thus the density normalisation parameter $f_0$ is dimensionless, and the radial bins are normalised by the corresponding overdensity radius. For a given halo mass definition, $M_\Delta$, we fit the profiles out to $R_{\Delta}$.  

\begin{figure}   
    \centering
        \includegraphics[width =\columnwidth]{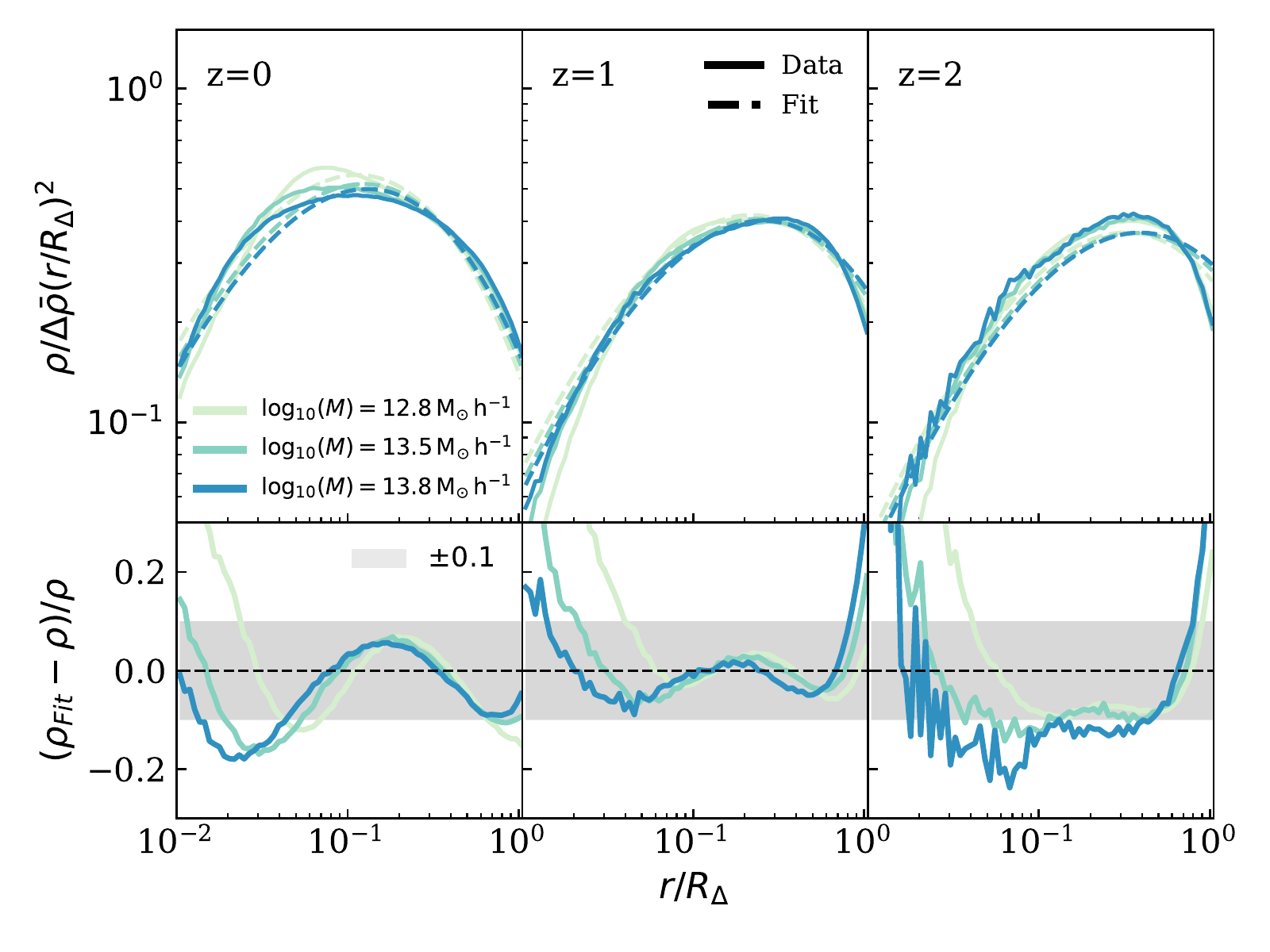}
        \caption{Stacked matter density (solid curves) and best-fit Einasto (dashed curves) profiles for the collisionless (\dm) case for three halo mass bins at three different redshifts ($z=[0,1,2]$) for the  $\Delta=200m$ mass definition.  In the top panels we show the profiles normalised by $\Delta\overline{\rho}$ and multiplied by $(r/R_{\Delta})^2$ to reduce the dynamic range. The bottom panels show the residuals of the best-fit Einasto profiles with respect to the stacked simulation profiles.  We have chosen the mass bins $\log_{10}(\rm M) = [12.8, 13.5, 13.8]$M$_{\odot}\,h^{-1}$ because they are well represented in all three redshifts.   The fitting functions reproduce the simulation profiles to typically $10\%$ accuracy over the full range of masses and redshifts and a radial range of $\approx[0.02-0.8] R_{\Delta}$.}
    \label{fig:stack_accuracy_dm}
\end{figure}{}

\begin{figure}   
    \centering
    \includegraphics[width =\columnwidth]{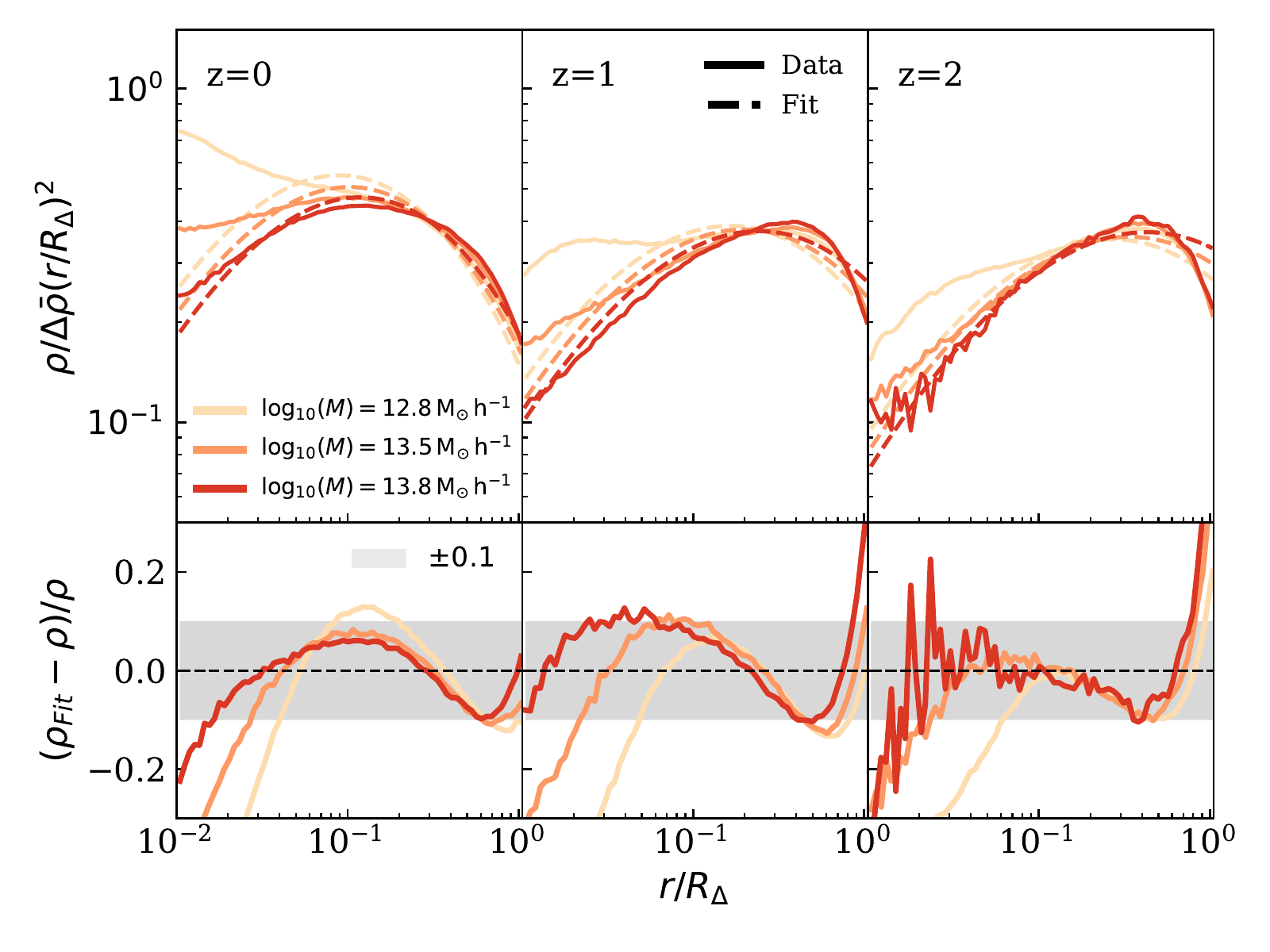}
    \caption{Same as Fig.~\ref{fig:stack_accuracy_dm} but for the hydrodynamical (\agn) case.  The strong deviations between the best-fit Einasto profile and the simulation profiles at low halo masses and small radii is due to the increasing importance of the central galaxy.  As we show later (see Fig.~\ref{fig:pk_ovr_stack_fit_results_agn}) this has a small but non-negligible effect on the non-linear power spectrum at small scales of $k \ga 5 [h/{\rm Mpc}]$.}
    \label{fig:stack_accuracy_agn}
\end{figure}{}

As an example of the profiles and the quality of the Einasto fits to them, in Figs.~\ref{fig:stack_accuracy_dm} and \ref{fig:stack_accuracy_agn} we present, at three different redshifts ($z=[0,1,2]$) and in three mass bins ($\log_{10}(M) = [12.8, 13.5, 13.8]\,$M$_{\odot}\,h^{-1}$), the stacked total matter density profiles for the \dm~and \agn~cases.  We show the case for a spherical overdensity of $\Delta  = 200m$, but find similar agreement for the other mass definitions. 
The density has been normalised by $200\overline{\rho}$ and multiplied by $(r/R_{\Delta=200m})^2$ in order to reduce the dynamic range of the plots. In the bottom panels, we present the residuals, defined as $(\rho_{\rm fit}-\rho_{\rm sim})/\rho_{\rm sim}$ with a shaded area that represents the $\pm\,10\%$ ($0.1$) agreement. For the \dm~case we have used different shades of blue while for the \agn~case different shades of red.  

The figures show that in both cases the generalised Einasto profiles can reproduce (typically within $10\%$) the radial trend of the profiles from $\approx[0.02-0.8]R_{\Delta}$ at all redshifts for the \dm~case.  There are some systematic features in the \agn~cases at radii below $\approx 0.07\,R_{\Delta}$. This increase in density is due to the stellar component\footnote{In principle one could include an additional component to the parametric model to better fit the inner regions.  Indeed, this would be recommended when modelling real data.  However, since we also evaluate the halo model using the tabulated profiles directly from the simulations, we can still assess the accuracy of the halo model without including such a component.  By comparing the tabulated and parametric versions, we can directly assess the impact of neglecting an additional component designed to better capture the central galaxy.} (and associated adiabatic contraction of the dark matter), particularly prominent in the innermost parts of relatively low-mass haloes.  This effect will also be slightly visible in the matter power spectrum analysis later for the differences with using the fitting profiles.    We also see that in the innermost radial bins at higher redshift the profiles are more noisy, which is just due to the expected lower abundance of very massive haloes at higher redshifts. 

\subsection{Halo mass function}
\label{hmf}

We now consider the halo mass function (HMF) from BAHAMAS as input for the halo model.  At a given redshift and for a given halo mass definition (spherical overdensity), we compute the halo mass function of FOF haloes using a bin width of $\log_{10}(M)=0.0625$, over a mass range $10^{11}-5\times10^{15}\,$M$_{\odot}\,h^{-1}$. To compute the mass function, $dn/dM$, we simply count the number of FOF haloes in a given bin and divide by the bin width and simulation comoving volume. 

\begin{figure}   
    \centering
    \includegraphics[width =\columnwidth]{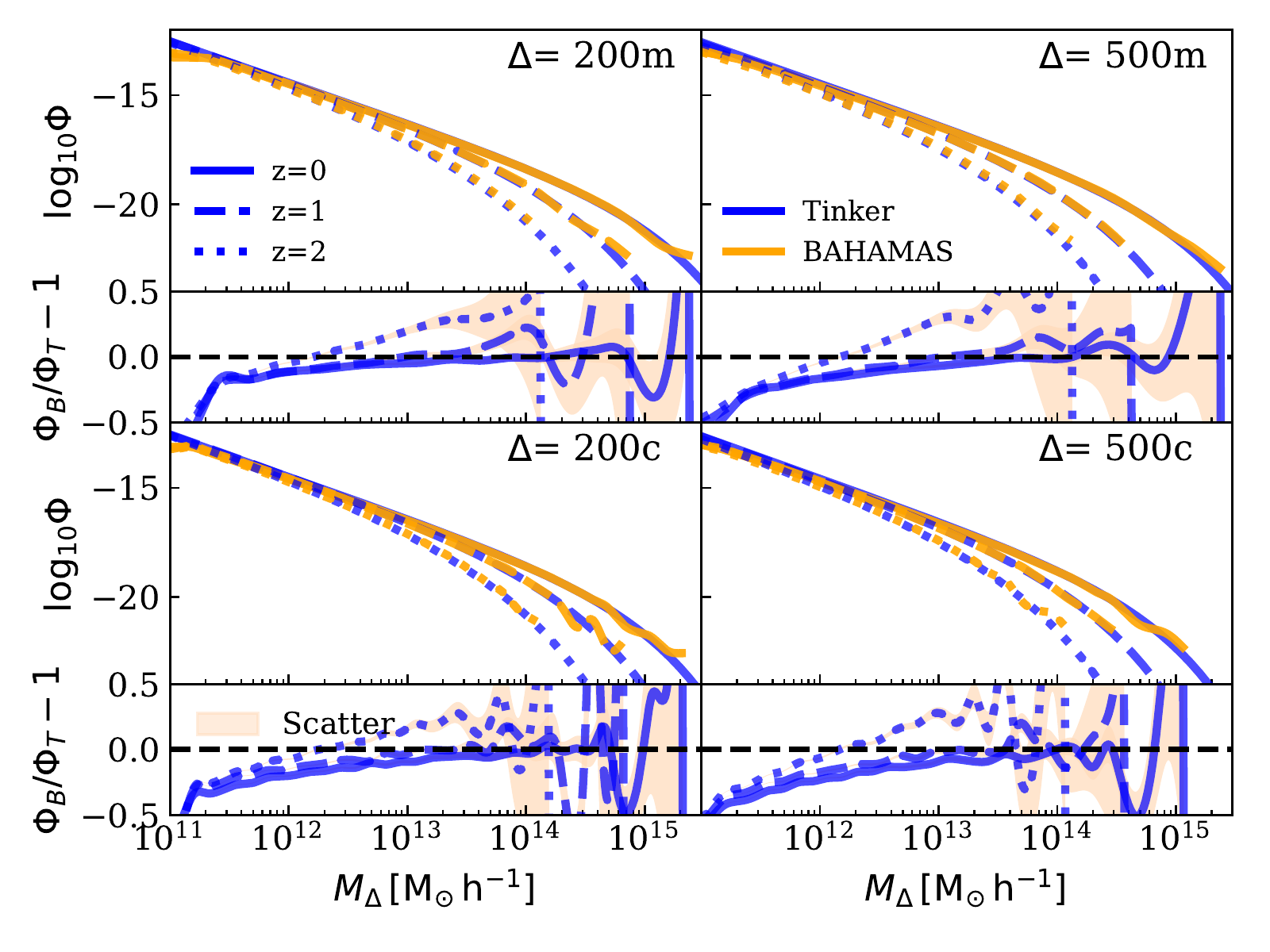}
    \caption{Halo mass function (HMF, $\Phi$) comparison between the BAHAMAS \dm~case ($\Phi_{B}$, orange curves) and the \citet{Tinker08} prediction ($\Phi_{T}$, blue curves) at three different redshifts (solid, dashed and dotted curves) for four different halo mass definition.  In the smaller subpanels residuals between the BAHAMAS HMF and the Tinker HMF are shown.  The shaded orange regions represent the Poisson errors for the BAHAMAS HMFs.  Differences between the Tinker and BAHAMAS mass functions are likely due to cosmic variance and Poisson uncertainties at the high-mass end and finite resolution and differences in how haloes are identified (FOF for BAHAMAS and spherical overdensity for Tinker) at the low-mass end (see text).  We examine how differences in the mass functions affect the resulting non-linear power spectrum in Fig.~\ref{fig:pk_ovr_stack_fit_results_dm}.}.
    \label{fig:hmf_dm_bahamas}
\end{figure}

We present in Fig.~\ref{fig:hmf_dm_bahamas} a comparison between the HMF from the BAHAMAS \dm~run with the \citet{Tinker08} prediction for the four different mass definitions at three different redshifts.  We have computed the Tinker HMF using the {\small Colossus Toolkit} \citep{diemercolossus} \footnote{\href{https://bdiemer.bitbucket.io/colossus/}{https://bdiemer.bitbucket.io/colossus/}}.

{}

We see that there is generally good agreement between the two independent mass functions. In the small panels below the main ones we present the residuals between the BAHAMAS and the Tinker HMFs for the three different redshifts (with the same lines). Small differences can be seen at high masses which are likely a result of cosmic variance and relatively poor statistics (Poisson errors) in the BAHAMAS volume.  Regardless of the origin of the differences, they should be taken into account evaluating the internal accuracy of the halo model.  For example, if through cosmic variance the BAHAMAS volume has somewhat more very massive clusters than expected on the basis of the Tinker HMF, this could also affect the overall non-linear $P(k)$ of the simulation.  Therefore, by using the actual HMF from BAHAMAS we can more accurately test the halo model formalism. 

At low masses ($\sim10^{11}\,$M$_{\odot}\,h^{-1}$), the BAHAMAS simulations predict a lower abundance of haloes compared to the Tinker expectation.  This is likely due to two effects: finite resolution of the BAHAMAS  simulations and differences in the way haloes are identified in BAHAMAS and \citet{Tinker08}.  There is a clear resolution effect at masses below $\approx 3\times10^{11}$ M$_\odot\,h^{-1}$, where the BAHAMAS HMF stops increasing with decreasing mass.  Here the simulations are approaching the 20 particle limit imposed on FOF groups.  At somewhat higher masses, there is still a deficit with respect to the Tinker prediction of $\approx10-20\%$.  This is likely due to differences in the way haloes are identified.  For BAHAMAS, haloes are identified with a FOF algorithm after which spherical overdensity masses are computed, whereas \citet{Tinker08} identify haloes using the spherical overdensity method and haloes are allowed to partially overlap.  Consequently, more intermediate/low mass haloes are identified in the vicinity of larger haloes using the spherical overdensity method, whereas a FOF algorithm will combine haloes into larger group in which they are sufficiently close to one another.  These differences have been previously discussed in the literature (e.g., \citealt{bocquet20}) so we will not discuss them further here.  However, such differences in the HMFs will propagate through the halo model and affect the predictions for $P(k)$.  We will show that the differences in the HMFs will impact the $P(k)$ predictions only slightly at low redshift, but play a relatively larger role at higher redshift ($z\approx2)$.  

\subsubsection{HMF baryon correction}
\label{hmf_baryon_correction}

In the HMF comparison presented above, we examined the \dm~run from BAHAMAS and compared it with the predictions of \citet{Tinker08}, who used a large suite of collisionless (dark matter-only) cosmological simulations to calibrate an approximately universal form (to $\sim10\%$ accuracy) for the HMF (see eqn.~\ref{eq:hmf}).  Thus, the comparison was a consistent one.  However, as several authors have shown previously, the halo profiles and HMFs can be affected by baryonic processes such as feedback from supernovae and AGN \citep{cui2014,Velliscig14,bocquet16, Mummery17, pfeifer2020, stafford2019}, with effects as large as $20\%$ in the HMF which is large enough to have a non-negligible impact on cosmological parameter inference \citep{cusworth2014,castro2020, debackere20}. 

To evaluate the impact of baryons on the HMF and how these translate to predictions of the halo model, we extend the formalism presented in \cite{Velliscig14} to correct the masses and HMFs.  We explore two different ways of accounting for baryons in the HMF.  In the first case, we can exploit the fact that the \dm~and \agn~runs have the same phases in the initial conditions, making it possible to match haloes between the two runs (using the unique particle IDs) on a halo-by-halo basis, as done previously in \citet{pfeifer2020} and \citet{stafford2019} when evaluating both the impact of baryons and cosmological extensions (dynamical dark energy and a running of scalar spectral index, respectively) on the HMF in BAHAMAS.  With this approach, one can directly determine how the halo mass has changed as a result of baryonic processes.  In the second approach, one can simply compare the HMFs of the \dm~and   \agn~runs, effectively computing the ratio of abundances in a given halo mass bin (e.g., \citealt{Velliscig14}).  By default we use the halo matching scheme to derive a HMF correction factor , but we have also explored an analysis using the HMF ratio method.  In short, while both approaches yield similar results, we find the halo matching scheme to be more accurate (less noisy).

\begin{figure}
    \centering
    \includegraphics[width =\columnwidth]{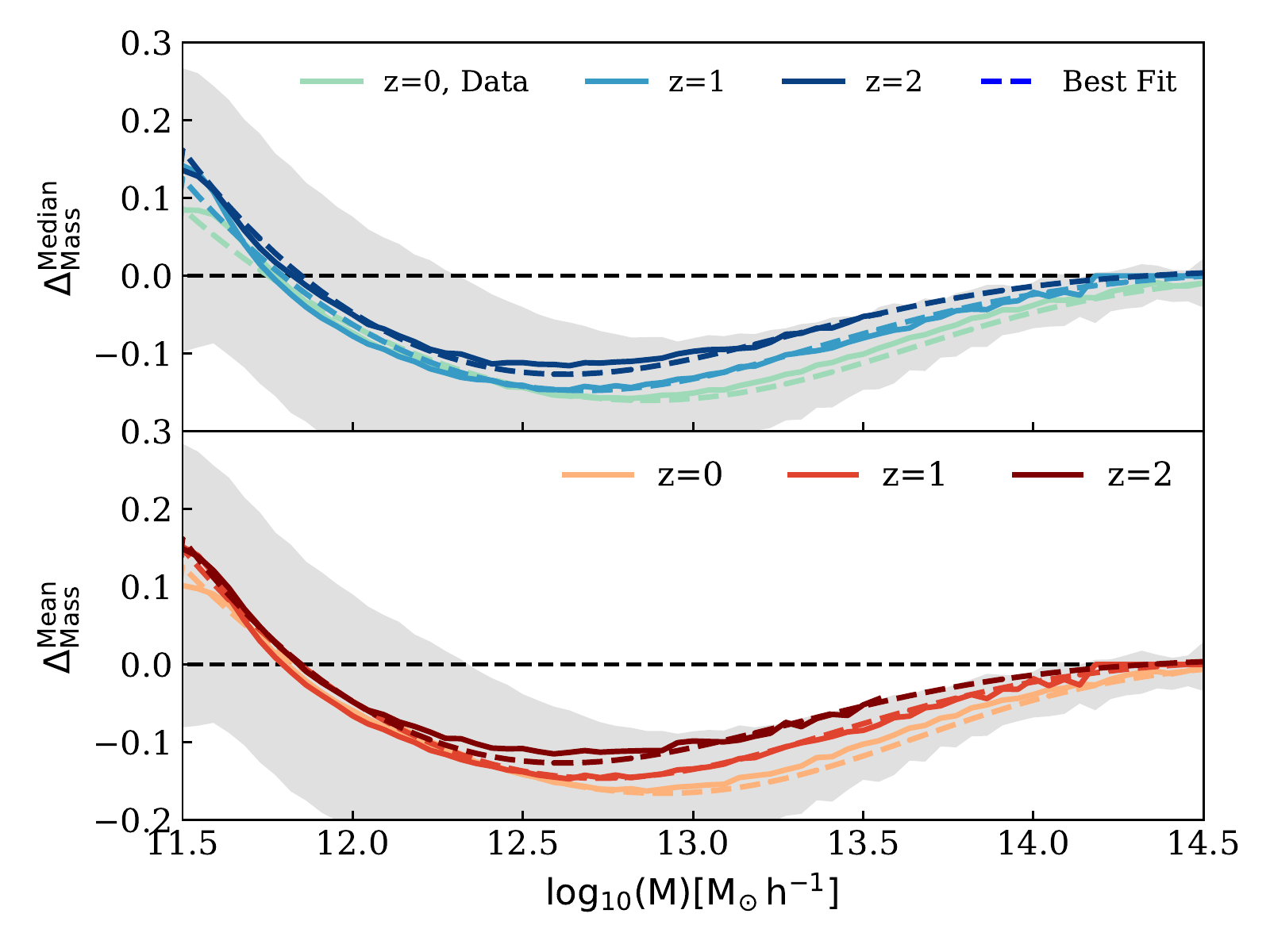}
    \caption{Fractional change in halo mass (eqn. \ref{eq:delta}) between the BAHAMAS \agn~and \dm simulations as a function of \dm~halo mass. Shown are the median (top panel) and mean (bottom panel) trends, along with the best-fit functions (eqn. \ref{velliscig_new}), for the $\Delta = 200m$ case at three different redshifts.  The scatter, shown in shaded grey area, represents the 68\% confidence region at $z=0$.}
    \label{fig:HMF_BAR}
\end{figure}

In Fig.~\ref{fig:HMF_BAR} we present a comparison between the mean and median values of $\Delta_{\rm mass}$ at three different redshifts for the $\Delta = 200m$ case, where $\Delta_{\rm mass}$ is the fractional difference in the halo mass between the \agn~and \dm~runs:
\begin{equation}
\Delta_{\rm mass} = \left(\frac{M_{\textit{AGN}} - M_{\textit{DM-Only}}}{M_{\textit{DM-Only}}}\right).
\label{eq:delta}
\end{equation}

Consistent with previous studies, we find that the halo masses are most strongly affected on the scale of galaxy groups, where AGN feedback is able to expel a large fraction of the baryons.  At higher masses (above a few $10^{14}$ M$_\odot\,h^{-1}$) the increased binding energy of the haloes prevents significant gas expulsion, while at lower masses ($\la 10^{12}$ M$_\odot\,h^{-1}$) AGN feedback is generally not yet active and stellar (supernova) feedback is not sufficiently energetic to eject a significant amount of baryons.

Following \citet{Velliscig14}, we model the change in halo mass (mass shift) due to baryons with the following functional form: 

\begin{eqnarray}
    \Delta_{\rm mass}(M_{\textit{DM-Only}}) & = & \frac{A}{\cosh[\log_{10}(M_{\textit{DM-Only}})]} \nonumber \\ 
    & + & \frac{B}{1+\exp\left[-\frac{\log_{10}(M_{\textit{DM-Only}}) - C}{D} \right]} \label{velliscig_new}
\end{eqnarray}

Note that in eqn.~\ref{velliscig_new} we have added a hyperbolic cosine term that allows the function to better reproduce the increase in $\Delta_{\rm mass}$ towards low halo masses.  In addition, to account for the redshift evolution of the halo mass shift, we allow the four parameters (A,B,C and D) to have power law redshift dependencies, e.g.,:
\begin{equation}
    A(z) = a_{0}(1+z)^{a_{z}}.
    \label{velliscig_par}
\end{equation}

In Fig.~\ref{fig:HMF_BAR} we see that the fitting functions can reproduce the halo mass shift for all mass bins ($>3\times10^{11}\,$ M$_{\odot}\,h^{-1}$) and at the three different redshifts shown.  Note that the lower limit of $3\times10^{11}\,$M$_{\odot}\,h^{-1}$ is dictated by the minimum number of matched most-bound particles (50) that we require to match haloes between two BAHAMAS runs.

Overall, the accuracy of the best-fit functions to the mean and the median values of $\Delta_{\rm mass}$ is better than $10\%$ in all mass bins and redshifts sampled.  In Table \ref{table:fitvalues_masscorrection} we present the best-fit parameter values for the mean and median versions of the mass shift $\Delta_{\rm mass}$.

\begin{table}
\centering
\caption{Best-fit parameters for the baryonic mass correction (eqn.~\ref{velliscig_new}) for the median and mean fractional changes in halo mass, $\Delta_{\rm mass}$. In Fig.~\ref{fig:HMF_BAR} we show the comparison between our best-fit models and the median and mean trends.  
}
\begin{tabular}{l | cc | cc} \hline
 & $\Delta_{\rm Mean}$ & & $\Delta_{\rm Median}$ & \\ 
P & $a_{0}$ & $a_{z}$ & $a_{0}$ & $a_{z}$\\
\hline
 A & $22291.3$ & $0.260$ & $19539.1$ & $0.417$\\
  B & $-0.327$ & $ 0.289$ & $-0.312$ & $0.414$\\
  C &$-13.574$ & $-0.033$ & $-13.552$ & $-0.037$\\
  D & $-0.395$ & $ 0.218$  & $-0.417$ & $0.216$\\
\hline
\end{tabular}
\label{table:fitvalues_masscorrection}
\end{table} 

As an aside, we find that the effects of baryons on the HMF (and presumably density profiles as well) are slightly cosmology dependent.  We have determined this by testing our model against a BAHAMAS \planck13 run that has a different universal baryon fraction ($f_{b}^{\rm Planck}=0.15433$) and we have found that the first parameter A depends upon this as 
\begin{equation}
A =A_{\rm WMAP9} \left(1+\frac{f_{b}}{f_{b}^{\rm WMAP9}}\right)^{-0.219},
\end{equation}
where $A_{\rm WMAP9}$ is the value presented in Table 
\ref{table:fitvalues_masscorrection}.  This cosmology correction works for both the mean and median $\Delta_{\rm mass}$ results. 

While the halo mass correction procedure derived above could be applied on a halo-by-halo basis to the BAHAMAS \dm~run to derive a baryon-corrected HMF, such a procedure would generally not be possible for published HMFs based on collisionless simulations, since the individual halo masses (halo catalogues) are generally not available.  Thus, we wish to derive a simple correction factor that can be applied to existing collisionless HMFs in the literature.

To do this, we first shift the halo mass bins from a collisionless HMF (in this case the BAHAMAS \dm~HMF) using the baryonic correction procedure above.  We use the mean correction function in Table \ref{table:fitvalues_masscorrection}.  This creates a new set of mass bins.  We next rescale the abundances ($\Phi$) by the relative ratio between the \dm~and \agn~mass bins as:


\begin{equation}
    \frac{dn}{dM_{\agn}} = \frac{dn}{dM_{\dm}} \frac{\left(M_{{\dm},i}-M_{{\dm},i-1} \equiv d M_{\dm}\right)}{\left(M_{\agn,i}-M_{\agn,i-1} \equiv d M_{\agn}\right)} \ \ \ , 
\label{new_phi}
\end{equation}
\noindent where $M_{\agn}$ is the corrected halo mass, derived using the (uncorrected) mass, $M_{\dm}$, and eqns.~\ref{eq:delta} and \ref{velliscig_new}, and $i$ refers to the i$^{\rm th}$ mass bin.  Essentially, this procedure works because the number density of haloes, $n$, does not change as a result of feedback/baryons, rather these result in a change in halo mass.  But a change in mass means that both the x-axis (halo mass) and the y-axis ($\phi \equiv dn/dM$, through the change in $dM$) change due to baryons\footnote{An underlying assumption of this procedure is that the \textit{rank ordering} of haloes by mass does not change through the inclusion of baryons.}.  With this procedure, we can correct existing HMFs derived from collisionless simulations for the presence of baryons.

\begin{figure}   
    \centering
        \includegraphics[width =\columnwidth]{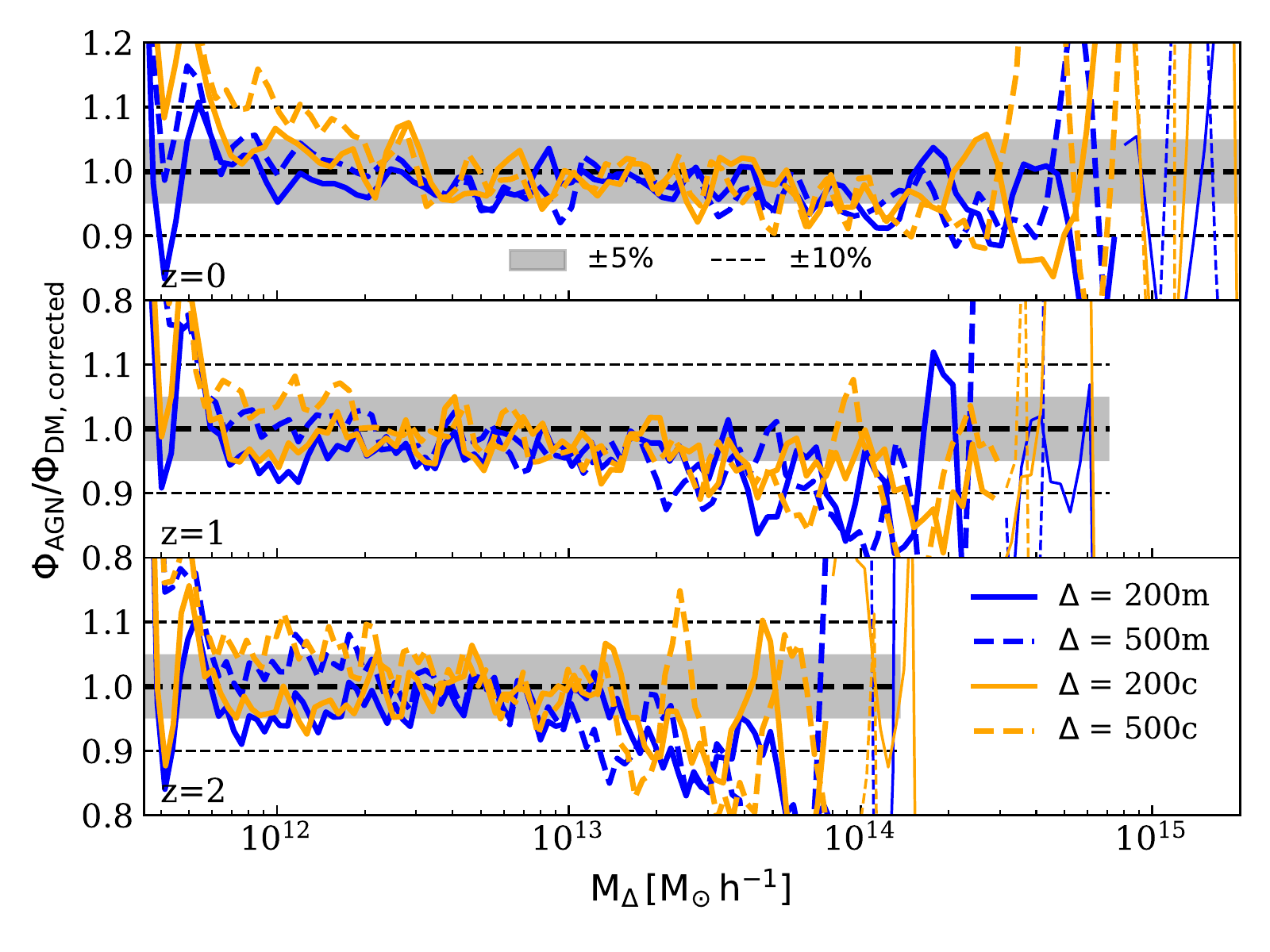} 
    \caption{Ratio of the BAHAMAS \agn~HMF over the baryon-corrected \dm~case.  The comparison is made for three different redshifts and all four mass definitions (orange for the critic mass definitions and blue for the mean-based definitions).  The baryonic correction helps recover the hydro simulation HMFs to typically better than $10\%$ in all cases, apart from at very high masses (due to large Poisson uncertainties in $\Phi$).}
    \label{fig:hmf_correction_new_method}
\end{figure}{}

In Fig.~\ref{fig:hmf_correction_new_method} we present the results of the application of this method to the BAHAMAS \dm~HMF and compare it with the actual BAHAMAS \agn~HMF.  We can see that the method is accurate to typically $5\%$.  Larger deviations are present at the very highest masses, which are due to poor sampling statistics.  

\begin{figure}   
    \centering
    \includegraphics[width =\columnwidth]{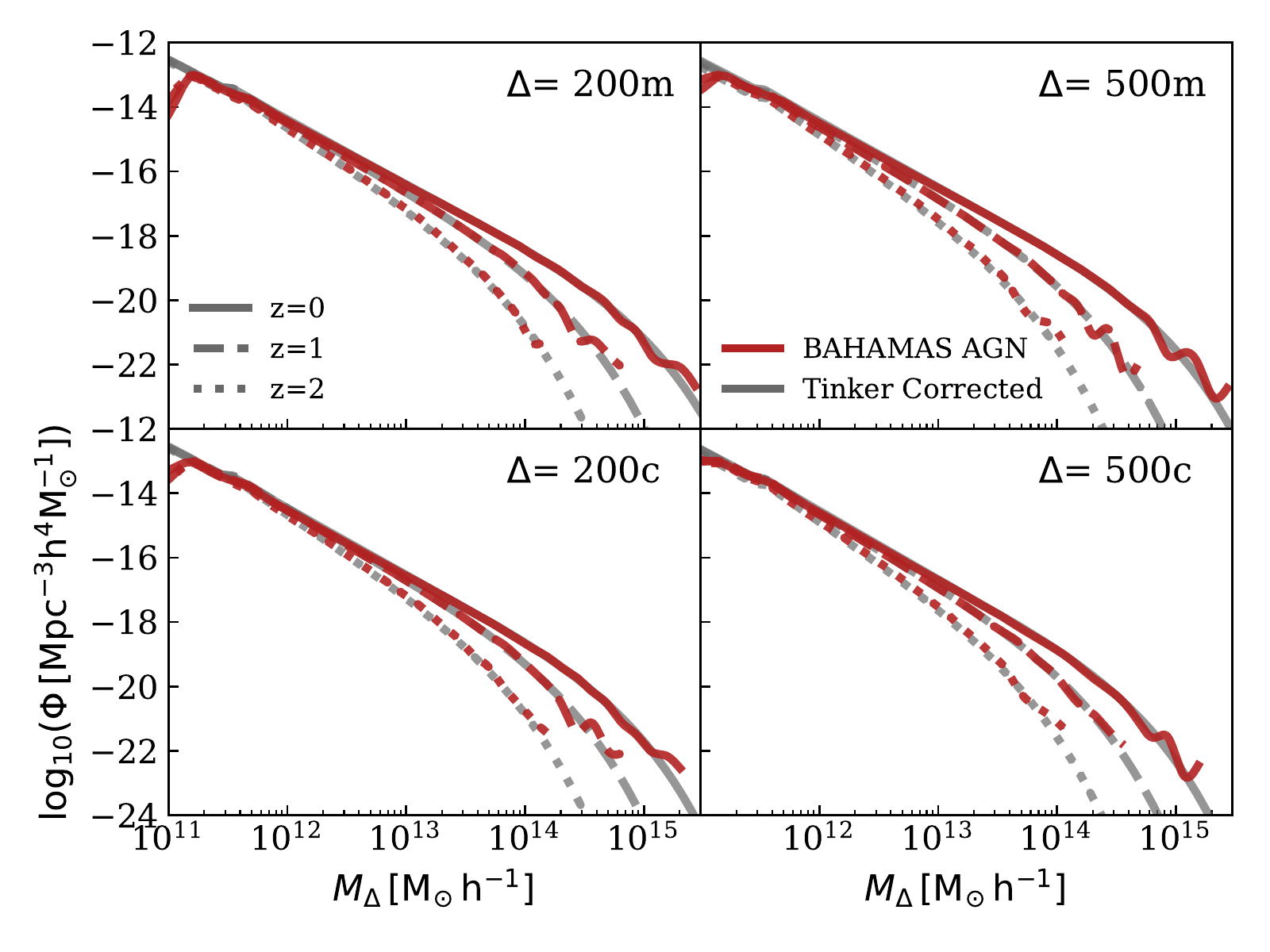} 
    \caption{Same as Fig.~\ref{fig:hmf_dm_bahamas} but for the \agn~case, with the Tinker HMF (grey curves) corrected using eqn.~\ref{velliscig_new} and compared with the BAHAMAS \agn~HMFs (red curves).}
    \label{fig:hmf_correction_tinker}
\end{figure}{}

In Fig.~\ref{fig:hmf_correction_tinker} we present a comparison between the baryon-corrected Tinker HMFs and the BAHAMAS \agn~HMFs, as we have done for the \dm~case in Fig.~\ref{fig:hmf_dm_bahamas}. We can see that the method applied also provides good results in this case. 

In this section we have explained how we extract the ingredients necessary from the BAHAMAS simulations to be able to evaluate the halo model.  We have shown the accuracy with which the Einasto fits reproduce the stacked density profiles computed from the simulations. We have also examined the differences in the HMFs, deriving a baryon correction factor that can be applied to HMFs from collisionless simulations.  Below we apply these quantities to calculate the non-linear power spectrum, $P(k)$, using the halo model and we compare this with the actual power spectrum measured from the BAHAMAS simulations.

\section{Matter power spectrum} 

\label{pk_exploration}
\subsection{Collisionless matter power spectrum}

In this section we present a comparison of the (BAHAMAS-informed) halo model predictions for the non-linear matter power spectrum alongside power spectrum predictions from the BAHAMAS simulations themselves. We also show linear theory prediction computed by \camb~ \citep{LewisChallinor} and the non-linear power spectrum from the (collisionless) \halofit~package \citep{takahashi12}.  Note that \halofit~provides a non-linear correction factor for the linear power spectrum, which \citet{takahashi12} have derived by fitting to a large suite of collisionless simulations spanning a wide range of cosmologies.

We begin by presenting the results for the collisionless (\dm) case.  As already discussed, we explore different versions of the halo model, where, for the density profiles we use either the tabulated profiles extracted directly from the simulations or a smooth parametric fit to them and for the halo mass function we use either the HMF directly from the simulations or forms from the literature (specifically \citealt{Tinker08}).  We also explore the impact of changing the halo mass definition, by varying the overdensity criteria used to define a halo's mass and its radial extent.

\begin{figure}   
    \centering
    \includegraphics[width =\columnwidth]{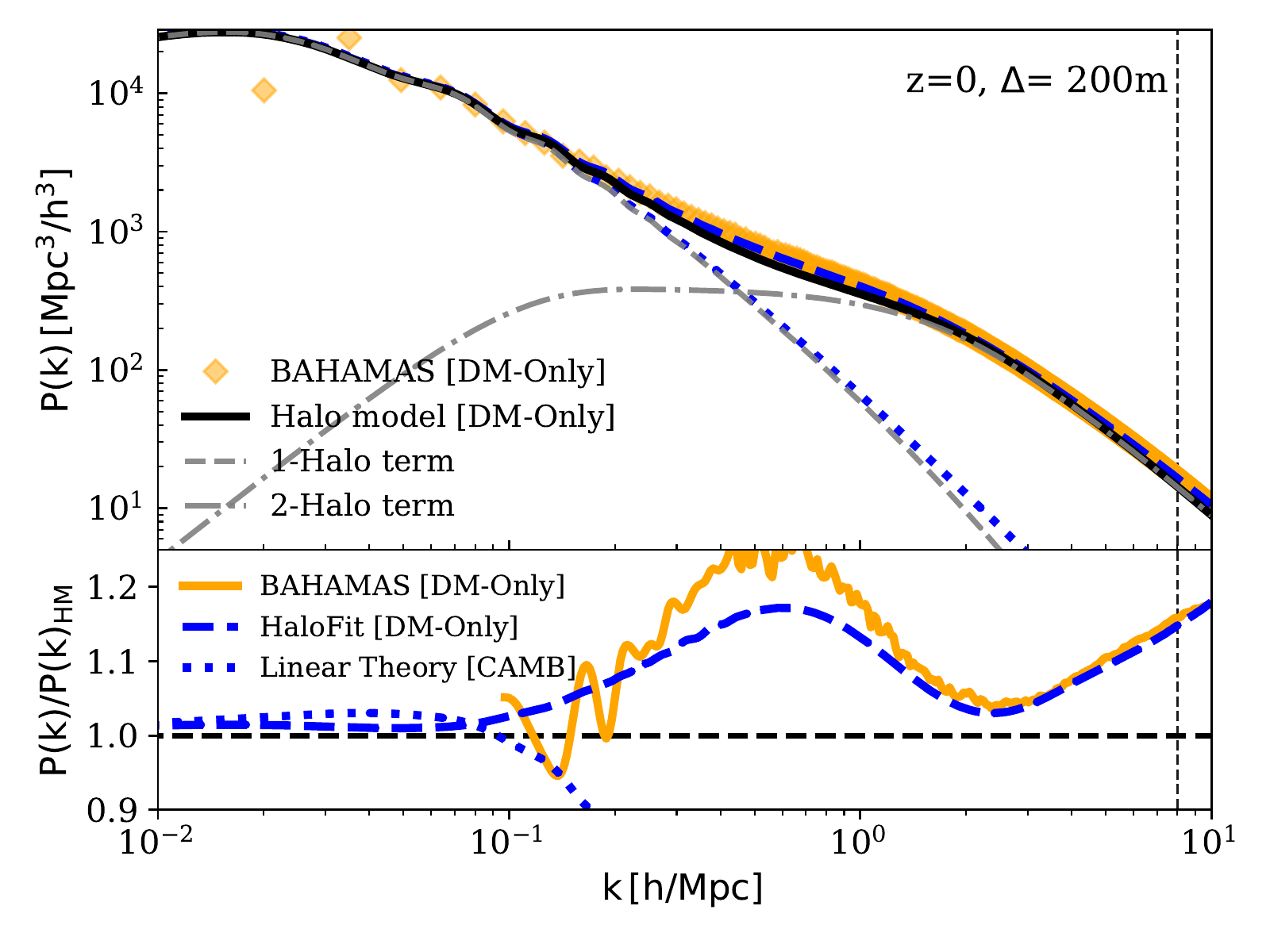}
    \caption{ {\textit Top:}  Matter power spectra comparison for the collisionless case.  The solid black curve represents the halo model computed with the stacked density profiles and the BAHAMAS HMF, while the dashed and dot-dashed grey curves represent the 1-halo and 2-halo terms separately.  The power spectrum from the BAHAMAS \dm~run is represented with orange symbols.  The dashed and dotted blue curves represent the non-linear power spectrum predicted by \halofit~and the linear matter power spectrum predicted by \camb.  The vertical dotted line represents 0.5 times the Nyquist frequency of the BAHAMAS simulation volume.  {\textit Bottom:}  Ratios with respect to the halo model.  On very large scales, the halo model reproduces linear theory to percent level accuracy, by construction.  On small scales, the 1-halo term dominates and reproduces the simulated (BAHAMAS) power spectrum to typically 5\% accuracy.  In the 1-halo/2-halo transition region, the halo model predicts up to 20\% less power than in the BAHAMAS simulations.}
    \label{fig:comparison_result_DMonly}
\end{figure}

In Fig.~\ref{fig:comparison_result_DMonly} we present the comparison between the halo model prediction (black solid line), its 1-halo and 2-halo terms (grey lines) and the BAHAMAS power spectrum (orange diamonds), as well as the predictions of linear theory and \halofit~(dotted and dashed blue curves, respectively).  Note that for this comparison, the halo model is computed using the tabulated mass density profiles (as opposed to Einasto fits to them) and the HMF directly from the BAHAMAS \dm~run.  In the bottom panel we present the ratio of the different cases with respect to the BAHAMAS-informed halo.  The BAHAMAS simulation power spectrum is computed using the software {\small NBodykit}\footnote{\href{https://nbodykit.readthedocs.io/en/latest/}{https://nbodykit.readthedocs.io/en/latest/}}\citep{hand2018}.

Qualitatively speaking, the halo model does capture the general trends of the simulation non-linear power spectrum well, including the shape of the power spectrum from the simulations (top panel).  For example, there is a strong increase in power with respect to linear theory on small scales, as expected.  Focusing on the bottom panel for a quantitative comparison, we can see that the halo model predictions match those of linear theory at large scales to percent level accuracy, which is by construction, after accounting for haloes that lie below the mass resolution limit of the simulations (see eqn.~\ref{small_haloes}).  The most challenging region is between $0.1\la k \la 2$ $h$/Mpc which corresponds to the transition region between the 1-halo and 2-halo terms.  Here the halo model's prediction can deviate from the simulations by up to $15\%$.  This is qualitatively consistent with previous findings (e.g., \citealt{giocoli10, massara14, Mead15,chen2020, voivodic2020}), although note that our test is more stringent due to the fact that we are using the same simulation to inform and then test the halo model. 

At small scales we see that the level of agreement improves again ($<5\%$ at $1 < k [h/\rm Mpc] < 4$) between the BAHAMAS-informed halo model predictions and the simulations and theoretical predictions.  However, the error increases again at still smaller scales.  While the error increases as the Nyquist frequency is approached\footnote{The Nyquist frequency is defined as $\nu_{y} = 2 \pi N_{\rm cell}/L_{\rm Box}$ where $N_{\rm cell}$ is the number of cells used in the Fourier transform (to the one-third power) when evaluating $P(k)$ and $L_{\rm Box}$ is the box size.} (see dotted vertical line), the fact that the \halofit~prediction is very similar to that of BAHAMAS suggests that the error is not solely due to aliasing effects in the simulation $P(k)$.  Further tests exploring the minimum radius and halo mass, as well as the the radial and mass binning strategies, in the halo model show the results to be numerically robust.  Plausible physical explanations for the deviation at very small scales include differences in the clustering of substructures compared to the smooth dark matter profile, asphericity of the mass distribution, and intrinsic scatter in the mass density profiles.

\begin{figure*}
\begin{multicols}{2}
    \includegraphics[width=\linewidth]{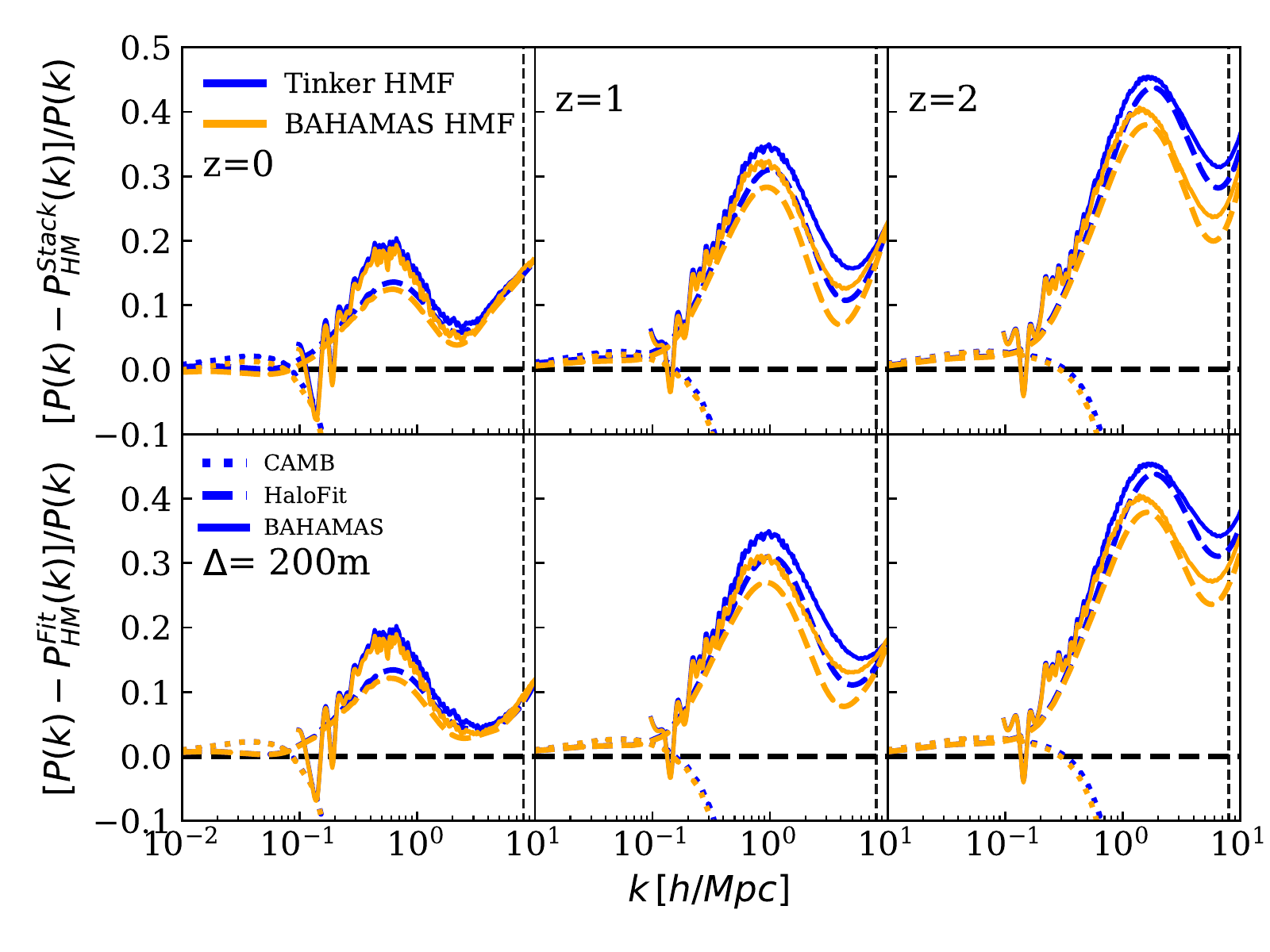}\par 
    \includegraphics[width=\linewidth]{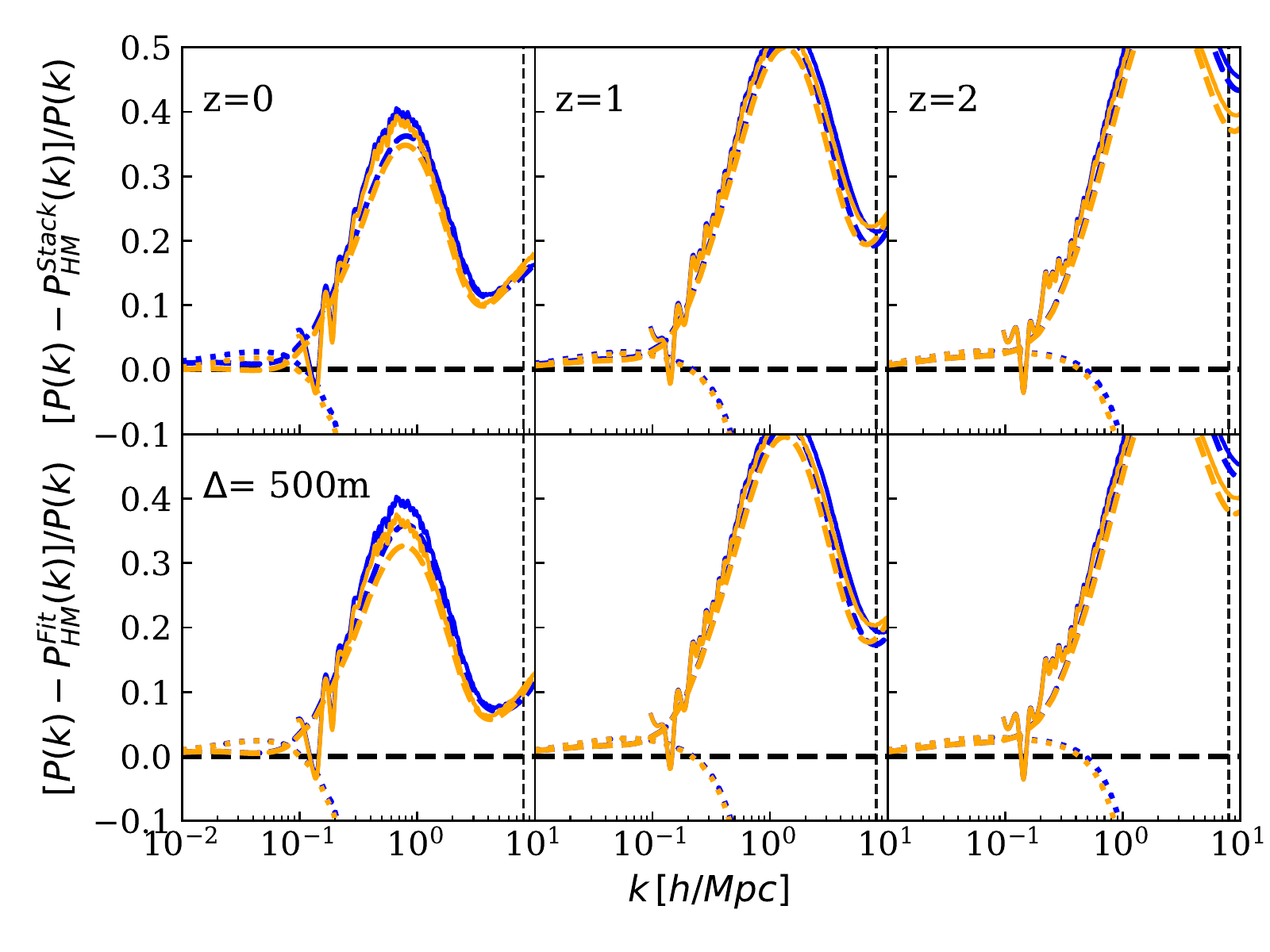}\par 
\end{multicols}
\begin{multicols}{2}
    \includegraphics[width=\linewidth]{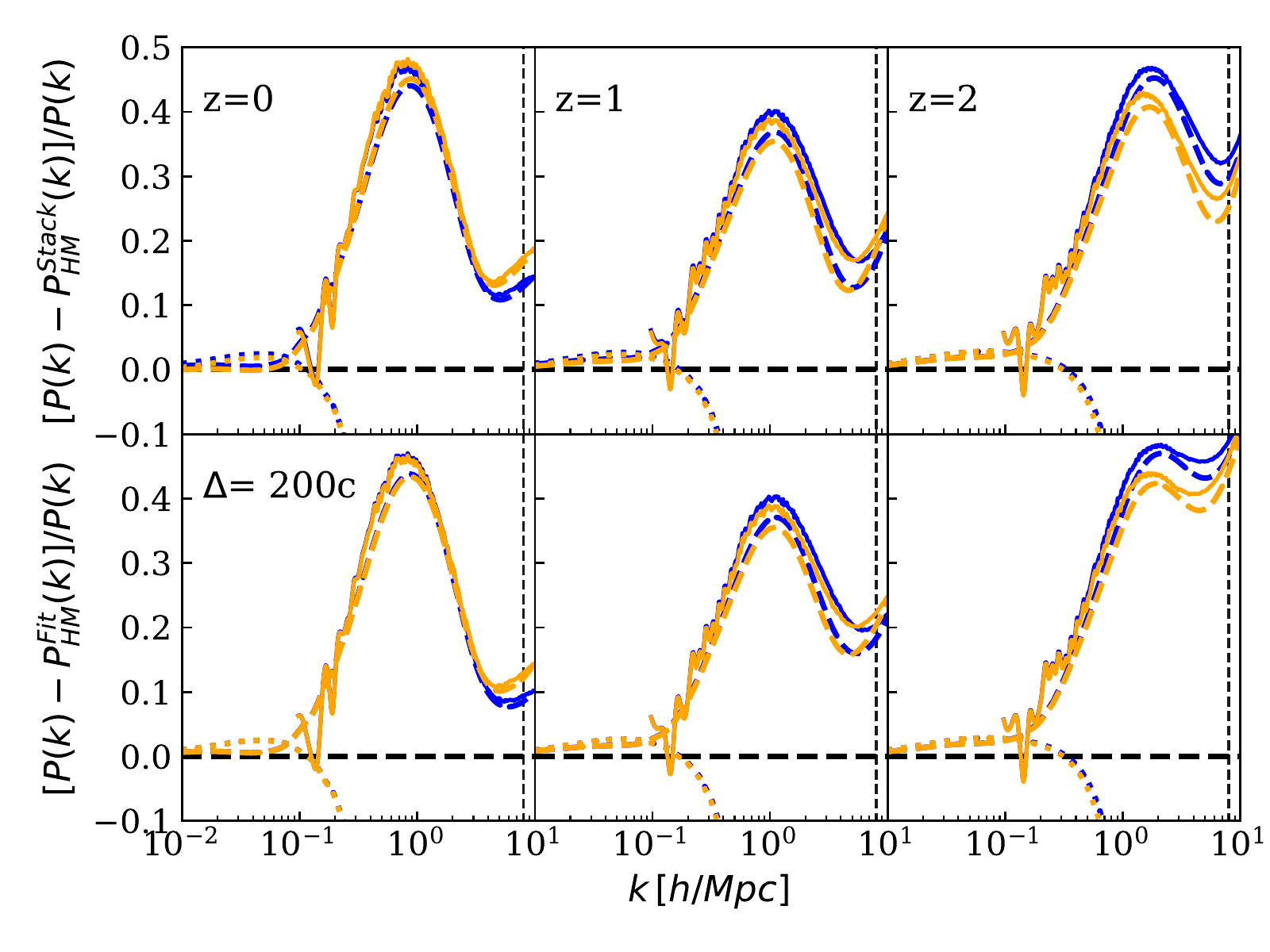}\par
    \includegraphics[width=\linewidth]{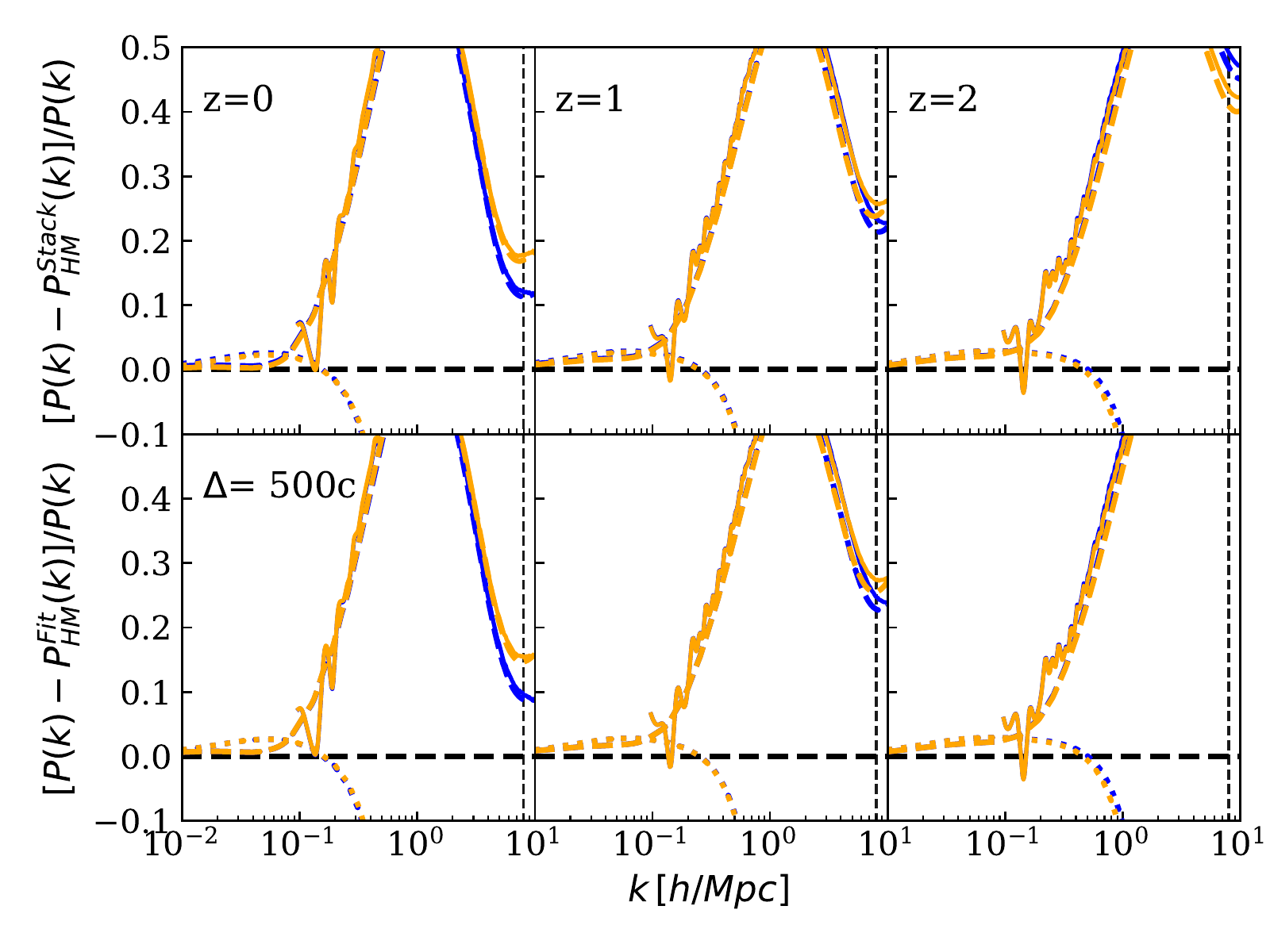}\par
\end{multicols}
\caption{Residual plots between the matter power spectrum and the halo model predictions at three different redshifts for the four different mass definition as follows: top-left $\Delta = 200m$, top-right $\Delta = 500m$, bottom-left $\Delta = 200c$ and bottom-right $\Delta = 500c$.  The top (bottom) row of panels in each plot set correspond to the case where the tabulated profiles directly from the simulations (Einasto fits to) are used in the halo model.  The orange (blue) curves correspond to the case where we use the BAHAMAS \dm~simulation (Tinker) HMF.  We compare against linear theory predictions using \camb~(dotted curves) and the non-linear \halofit~ prediction (dashed curves) and BAHAMAS \dm~simulation (solid curves). The vertical dotted line represents 0.5 times the Nyquist frequency of the BAHAMAS simulation volume.  All models recover the large-scale limit by construction while the 1-halo/2-halo transition region is (at best, corresponding to the $\Delta = 200m$ case) recovered to 10-15\% accuracy.  In general, the accuracy decreases with decreasing radial extent of the haloes (due to changing halo mass definition, see text) and increasing redshift.}
\label{fig:pk_ovr_stack_fit_results_dm}
\end{figure*}

In Fig.~\ref{fig:pk_ovr_stack_fit_results_dm} we explore the effects of changing the halo mass definition, the profiles (tabulated vs. parametric fit), and the HMF (BAHAMAS vs.~\citealt{Tinker08}) at a number of different redshifts.   There are four sets of plots, corresponding to the four halo mass definitions that we explore ($\Delta = 200m$, $500m$ and $\Delta =200c$, $500c$).  The top row of panels in each plot set correspond to the case where the tabulated profiles directly from the simulations are used in the halo model, whereas the bottom row of panels use the Einasto fit to the density profiles.  The blue curves correspond to the case where we use the Tinker HMF, whereas the orange curves used the BAHAMAS \dm~simulation HMF.  Note that here we present residuals, defined as $[P(k)-P^{\rm HM}(k)]/P(k)$ (where $P^{\rm HM}(k)$ is the halo model prediction), whereas in the bottom panel of Fig.~\ref{fig:comparison_result_DMonly} we showed a simple ratio.

We focus first on the top left set of plots, corresponding to a spherical overdensity case of $\Delta = 200m$.  Scanning from left to right, it is clear to see the halo model increasingly struggles to capture the 1-halo/2-halo transition region with increasing redshift.  This is true regardless of which mass function we use (BAHAMAS or Tinker) or whether we use tabulated or fitted density profiles (top vs. bottom rows).  Interestingly, examining the other spherical overdensity cases (see the other three sets of plots in Fig.~\ref{fig:pk_ovr_stack_fit_results_dm}), it appears that when the overdensity criteria are defined with respect to the mean background density, the precision of the model worsens with increasing redshift while the accuracy is mostly independent of redshift when the critical density is used.  The fact that there is a relation between the accuracy of the halo model and the mass definition was also hinted at in \citet{mead2020}, where they identified differences between using $\Delta = 200m$ and $\Delta = 200c$.

Comparing the top and bottom rows of the top left set of plots, there are no significant differences in the ability of the halo model to recover the simulation $P(k)$.  This implies that the Einasto form we have used reproduces the simulated matter density profiles sufficiently well for the purposes of predicting $P(k)$, since the result does not change when we use tabulated profiles directly (top row) vs. the Einasto fitting function (bottom row).

Comparing the solid orange (BAHAMAS HMF) and solid blue (Tinker HMF) curves, we see that using the actual BAHAMAS simulation HMF results in an improved agreement between the halo model and the simulation $P(k)$, particularly at higher redshifts.  Thus, the halo model is more accurate than what might have been inferred using a generic halo mass function to test it.

Scanning between the four sets of plots, another trend that is clearly visible is that changing the halo mass definition has a significant impact on the accuracy of the halo model with respect to the simulations.  The change in the halo mass itself is not what is driving this trend: since we essentially integrate over \textit{all} haloes, how we label their masses should not matter.  However, by changing the halo mass definition, we are also changing the radial extent (size) of a halo (given the spherical overdensity definition) and this clearly will impact where the 1-halo and 2-halo terms intersect, due to the change in the extent of the 1-halo term.  These results are consistent with the findings of \citet{vanDaalen2015}, who showed the importance of the radial selection of particles on the resulting power spectrum of cosmological simulations (see figure 3 of that study).  

We find that the larger the radial extent of the halo (noting that at $z=0$, $R_{200m}$ is the largest and $R_{500c}$ is the smallest) the better the halo model is able to capture the 1-halo/2-halo transition region in the simulations.  This suggests that one way to help further improve the halo model is to radially extend the 1-halo term, as has also been suggested recently by \citet{garcia2021}.  For example, even if the halo mass function and bias are defined with respect to some standard choice of overdensity (e.g., $\Delta = 200c,m$), the profiles could, for example, be extended to several times the corresponding spherical overdensity radius, with the optimum extent determined by fitting to the simulation $P(k)$.  However, whether such an approach is strongly cosmology dependent is unclear.  Alternatively, it may be possible to adopt a consistent mass and radius definition but simply lower the overdensity value (e.g., $\Delta=100$) or adopt an alternative physical mass/radius scale such as the `splashback' radius (e.g., \citealt{diemer15,diemer20}).  Finally, including an accurate treatment of non-linear bias should also help to better recover the transition region \citep{meadverde}.  We will examine these possibilities in future work.

\subsection{Matter power spectrum including baryon physics} \label{AGN_PK}

\begin{figure}   
    \centering
                \includegraphics[width =\columnwidth]{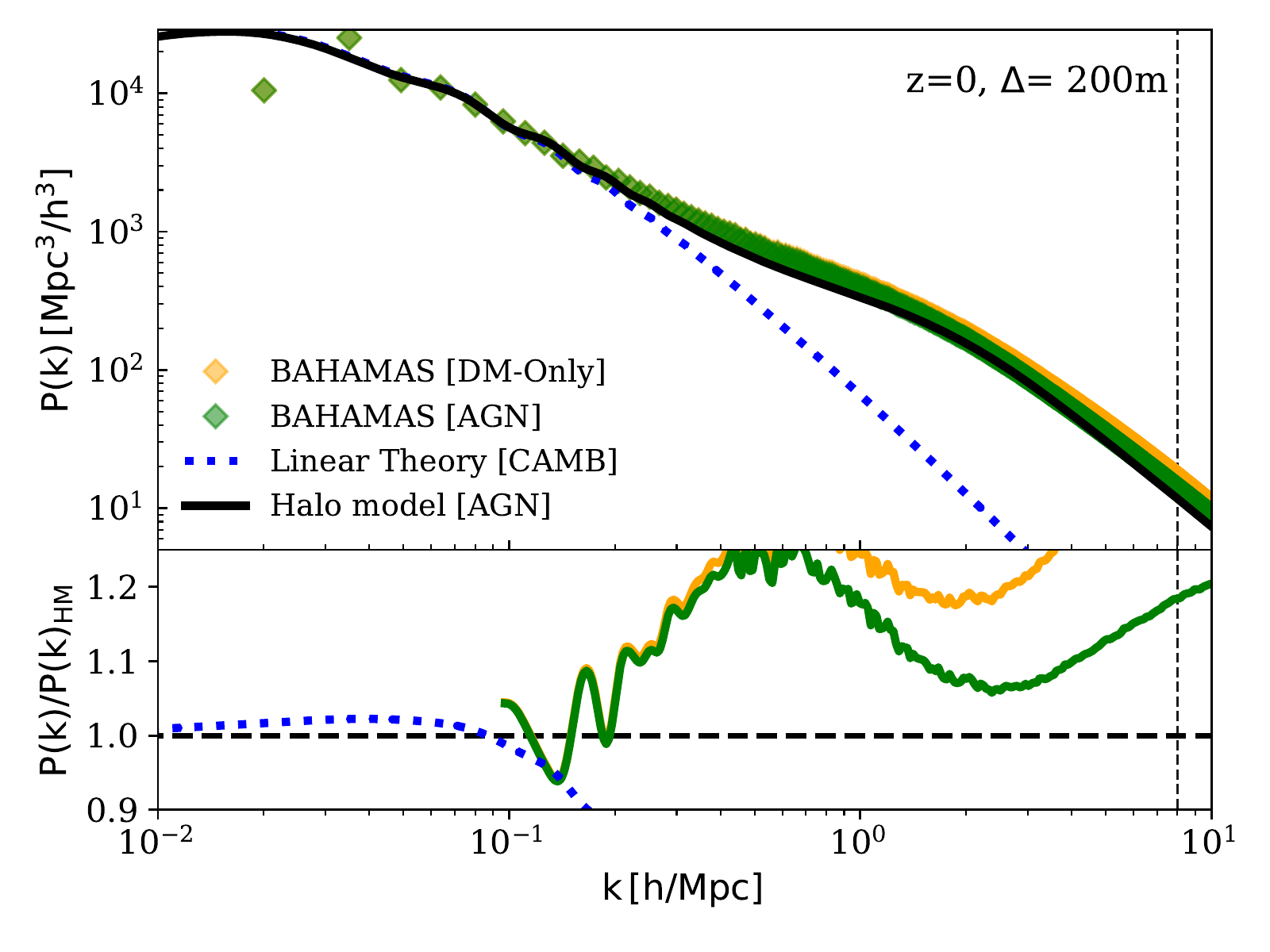} 
    \caption{Same as Fig.~\ref{fig:comparison_result_DMonly} but for the hydrodynamical (\agn) case. The BAHAMAS \agn~matter power spectrum is represented by green symbols.  The bottom panel we show the ratios with respect to the baryonic version of the halo model (i.e., using the density profiles and HMF from the BAHAMAS \agn~run).  The level of agreement between the baryonic halo model and the hydrodynamical simulations is similar to that seen in the comparison of the collisionless halo model and collisionless simulations in Fig.~\ref{fig:comparison_result_DMonly}.}
    \label{fig:comparison_result_AGN}
\end{figure}{}

In Fig.~\ref{fig:comparison_result_AGN} we present an analogous plot as in Fig.~\ref{fig:comparison_result_DMonly}, where the green diamonds represent the power spectrum from the BAHAMAS \agn~run and in the bottom panel the various ratios are now with respect to the baryon version of the halo model.  Note that the baryon version of the halo model corresponds to either using tabulated profiles directly from the \agn~run or an Einasto fit to them, as well as using either the BAHAMAS \agn~HMF or a Tinker HMF with a baryon correction applied.  For Fig.~\ref{fig:comparison_result_AGN} we use the tabulated profiles and HMF from the BAHAMAS \agn~run.

As in the case of the \dm~version, our baryon halo model prescription recovers the linear regime ($k<0.1h/$Mpc) to better than percent level accuracy, by construction.  Consistent with the collisionless comparison, the agreement is worst at the 1-halo/2-halo transition region, deviating from the simulation prediction by up to 20\%. The agreement improves again at smaller scales, though still deviates by $\approx$10\%.  

\begin{figure*}
\begin{multicols}{2}
    \includegraphics[width=\linewidth]{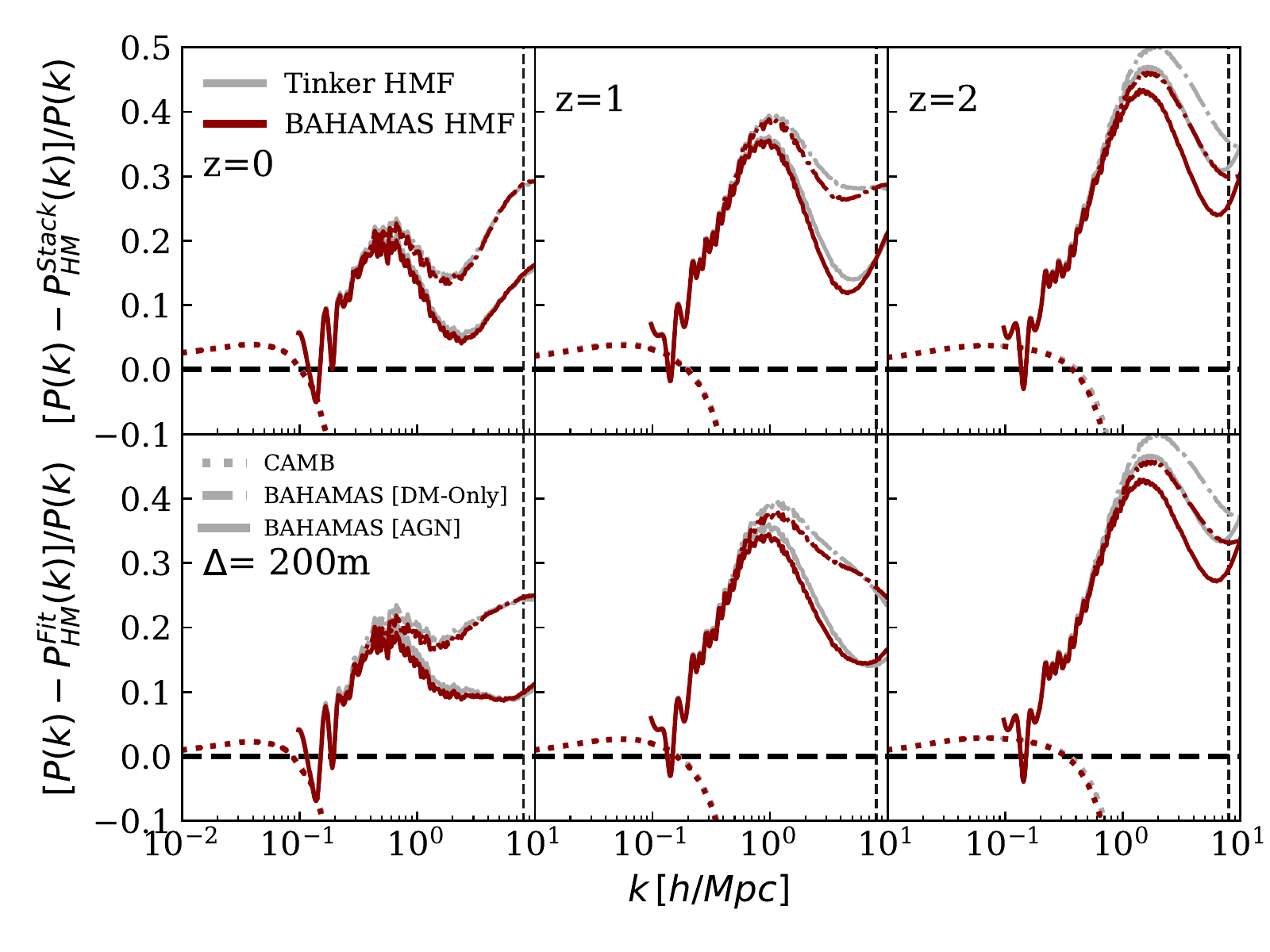}\par 
    \includegraphics[width=\linewidth]{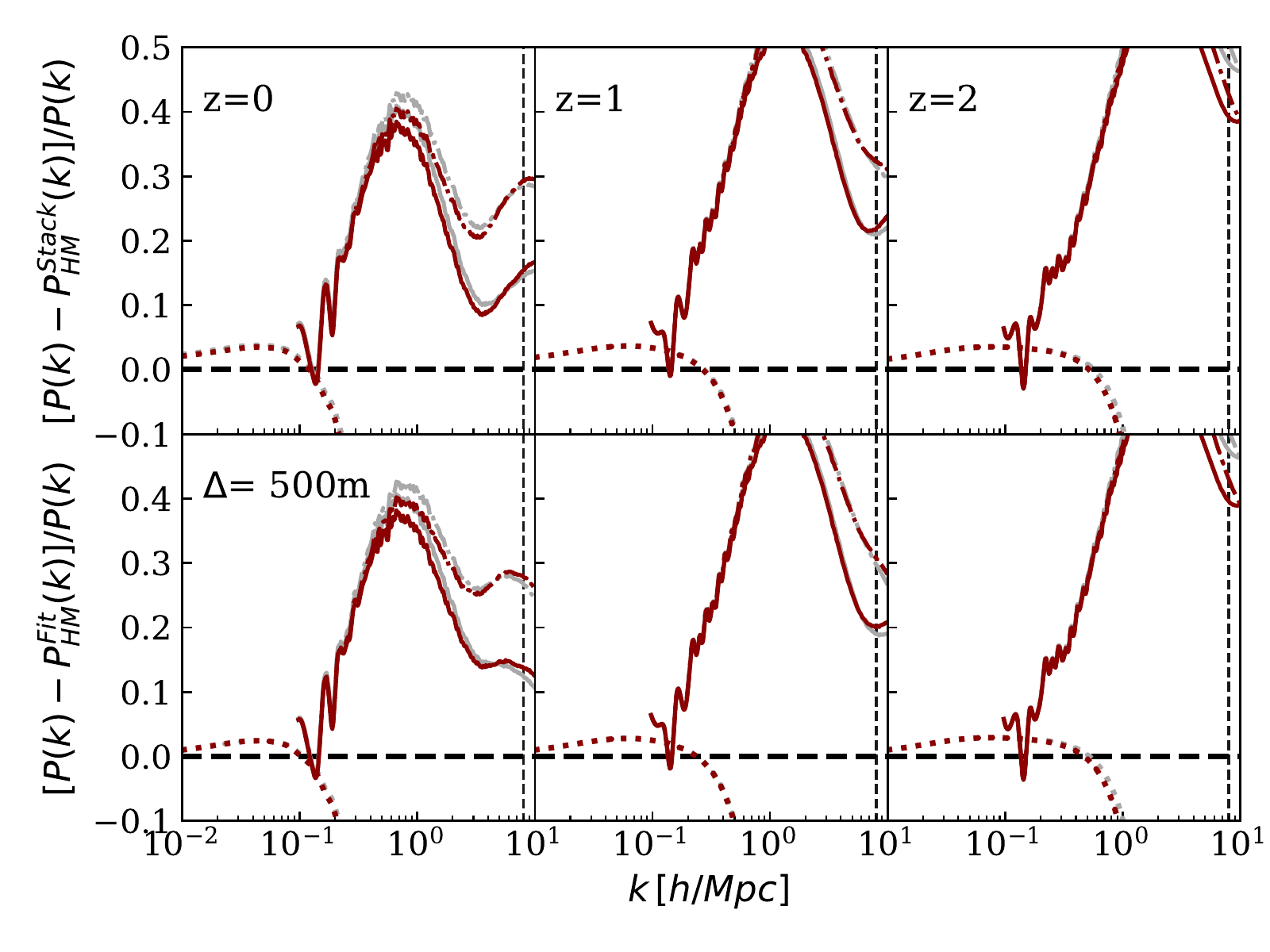}\par 
\end{multicols}
\begin{multicols}{2}
    \includegraphics[width=\linewidth]{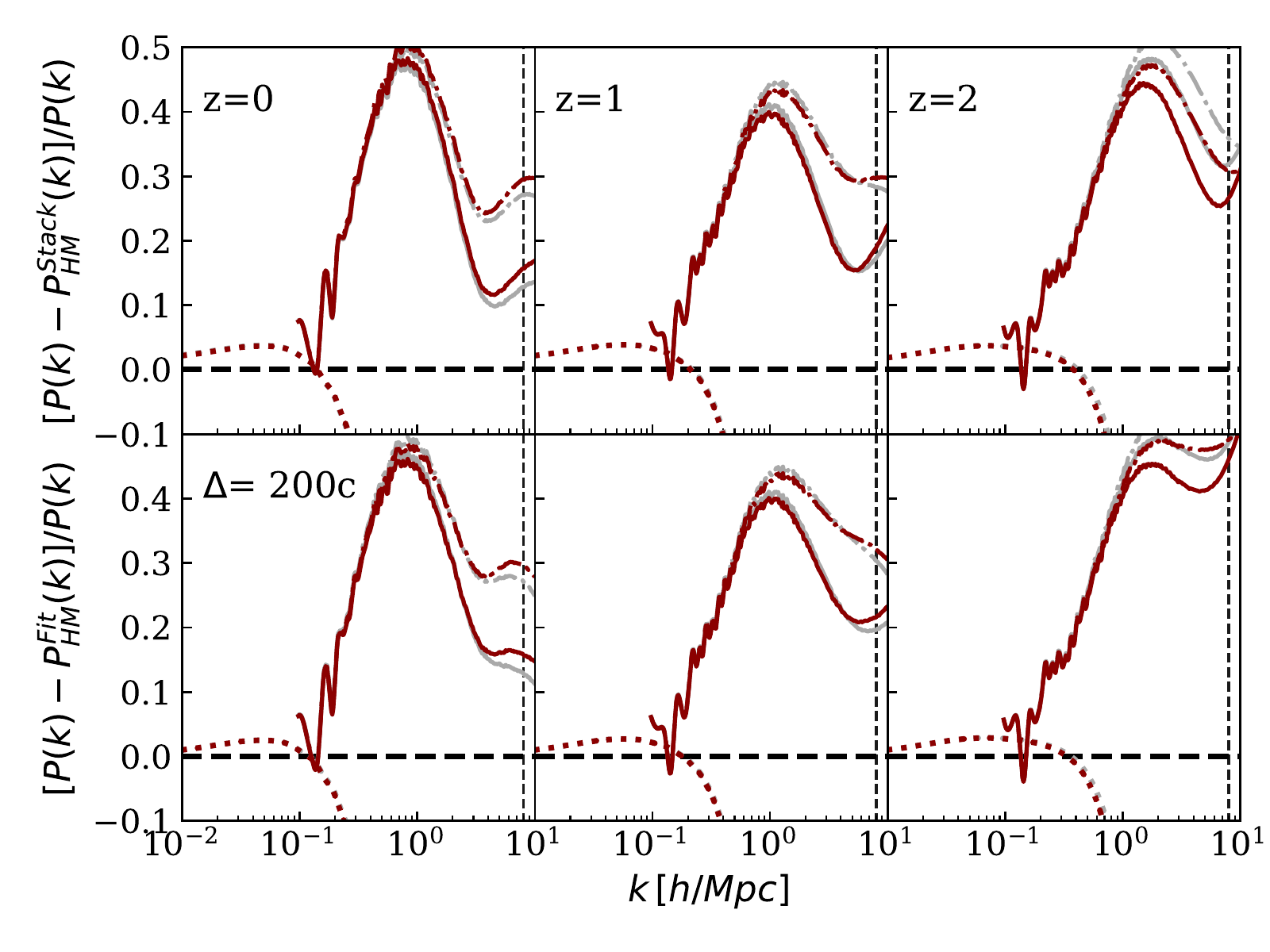}\par
    \includegraphics[width=\linewidth]{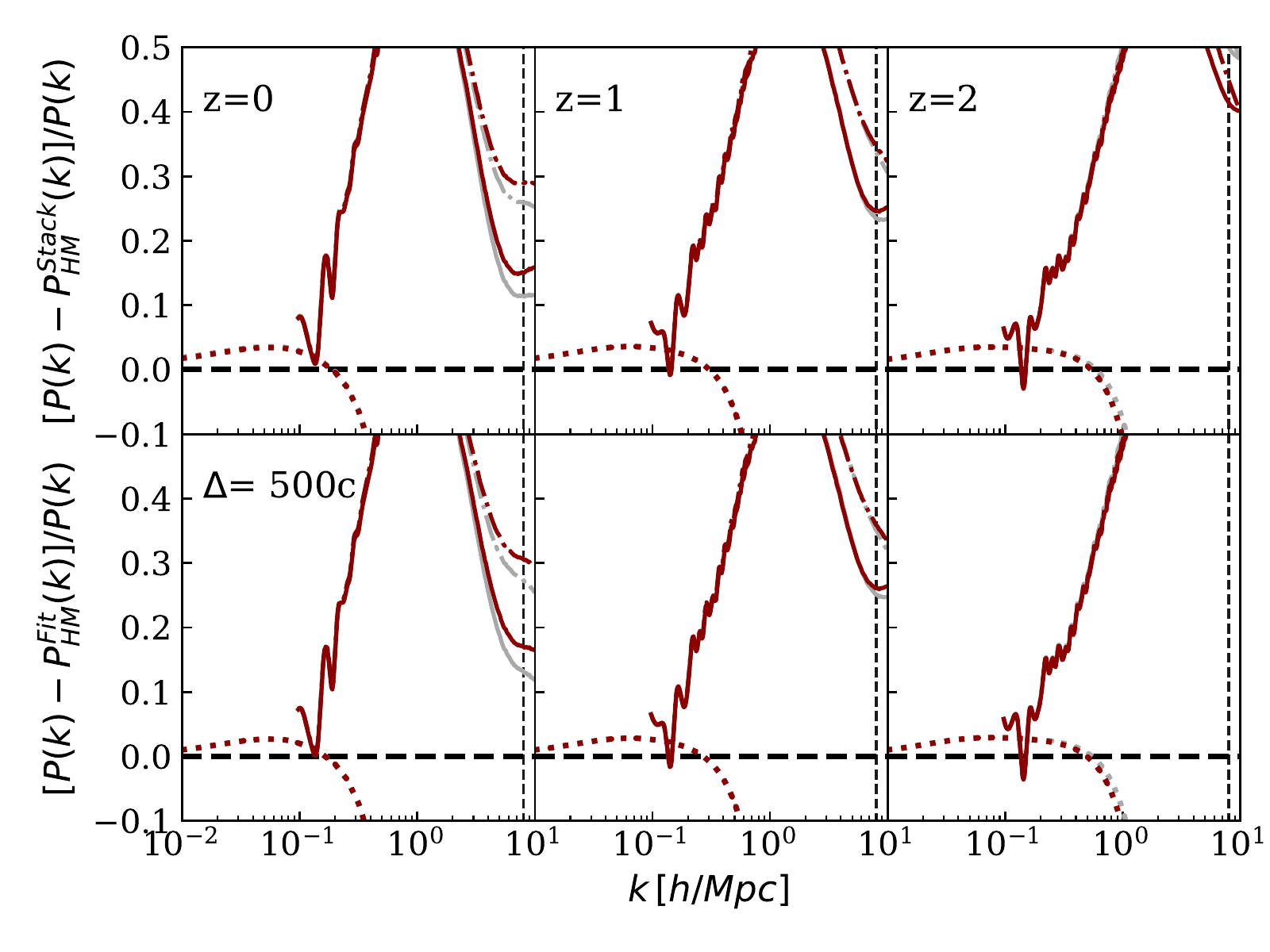}\par
\end{multicols}
\caption{Residuals plots between the baryonic matter power spectrum and the halo model predictions at three different redshifts for the four different mass definition following the same format as the \dm~companion plot in Fig.~\ref{fig:pk_ovr_stack_fit_results_dm}.  We also show, instead of the \halofit~prediction, the difference between the baryonic halo model compared the BAHAMAS \dm~$P(k)$ (dashed curve).  The overall trends and level of agreement are very similar to those found for the \dm~case, though there are differences in detail (see text).} 
\label{fig:pk_ovr_stack_fit_results_agn}
\end{figure*}

In Fig.~\ref{fig:pk_ovr_stack_fit_results_agn} we show the residuals plots in the same way we have presented for the \dm~case, with the stacked and fitted density profiles and two different forms for the HMFs, at different redshifts ($z=[0,1,2]$), and for the four different halo mass definitions.
The grey curves correspond to the cases using the baryon-corrected Tinker HMF and the red curves correspond to the cases using the BAHAMAS \agn~HMF. Overall, we find very similar trends to those presented in Fig.~\ref{fig:pk_ovr_stack_fit_results_dm} for the \dm~case.  Specifically, the mass definition that works best is again $\Delta = 200m$, which can recover the 1-halo/2-halo transition region to $20\%$ at $z=0$ and $\approx35\%$ at $z=1$. The 1-halo region ($k>2$ $h$/Mpc) is generally recovered to $10\%$ at $z=0$ independent of the choice of halo mass definition, HMF, and non-parametric vs.~parametric profiles.  At higher redshifts, the discrepancy with respect to the simulations increases for both the 1-halo/2-halo transition region and in the deep non-linear (1-halo) region.. 

Upon closer inspection, it is apparent in some cases that there is a difference at very small scales (high $k$ values) between the accuracy of the halo model when using either the parametric (Einasto) or tabulated mass density profiles.  For example, at $z=0$ in either the $\Delta = 200m$ or $\Delta = 200c$ cases, the residuals increase towards smaller scales when using the tabulated profiles, whereas for the parametric case they are approximately independent of $k$ scale.  We attribute this difference in behaviour to the inability of the Einasto form to fully capture the behaviour of the density profiles at small scales, due to the increasing importance of the central galaxy (see Fig.~\ref{fig:stack_accuracy_agn}).  Thus, in this case, using the more accurate tabulated density profiles demonstrates that the halo model is actually less accurate in reproducing the non-linear power spectrum on small scales.  

Overall, therefore, the trends in the accuracy of the baryon version of the halo model are very similar to those for the collisionless version, when the models are compared to the hydrodynamical and collisionless BAHAMAS simulations respectively.  In particular, we find that the absolute accuracy is worse at the 1-halo/2-halo transition and typically worsens at higher redshifts when the halo mass definition is defined with respect to the mean background density.  The choice of halo mass definition is also important.  Given that the trends are very similar between the baryon and collisionless cases, it raises the interesting question of whether the halo model would actually be better suited at predicting the \textit{ratio} (or suppression) of the matter power spectrum due to baryons, as opposed to predicting the absolute $P(k)$.  We explore this possibility below.

\subsection{Matter power spectrum suppression}
\label{Baryonic_matter_suppression}

We now explore the halo model predictions for the matter power spectrum suppression, sometimes also referred to as the ``suppression factor''.  There are many recent studies of the suppression factor using cosmological hydrodynamical simulations in the literature (e.g., \citealt{vandaalen11, vanDaalen2020,schneider2015,schneider19,chisari19,debackere2019}).
Here we explore the accuracy with which the halo model can recover the suppression of the matter power spectrum in the BAHAMAS simulations.  

\begin{figure*}
\begin{multicols}{2}
    \includegraphics[width=\linewidth]{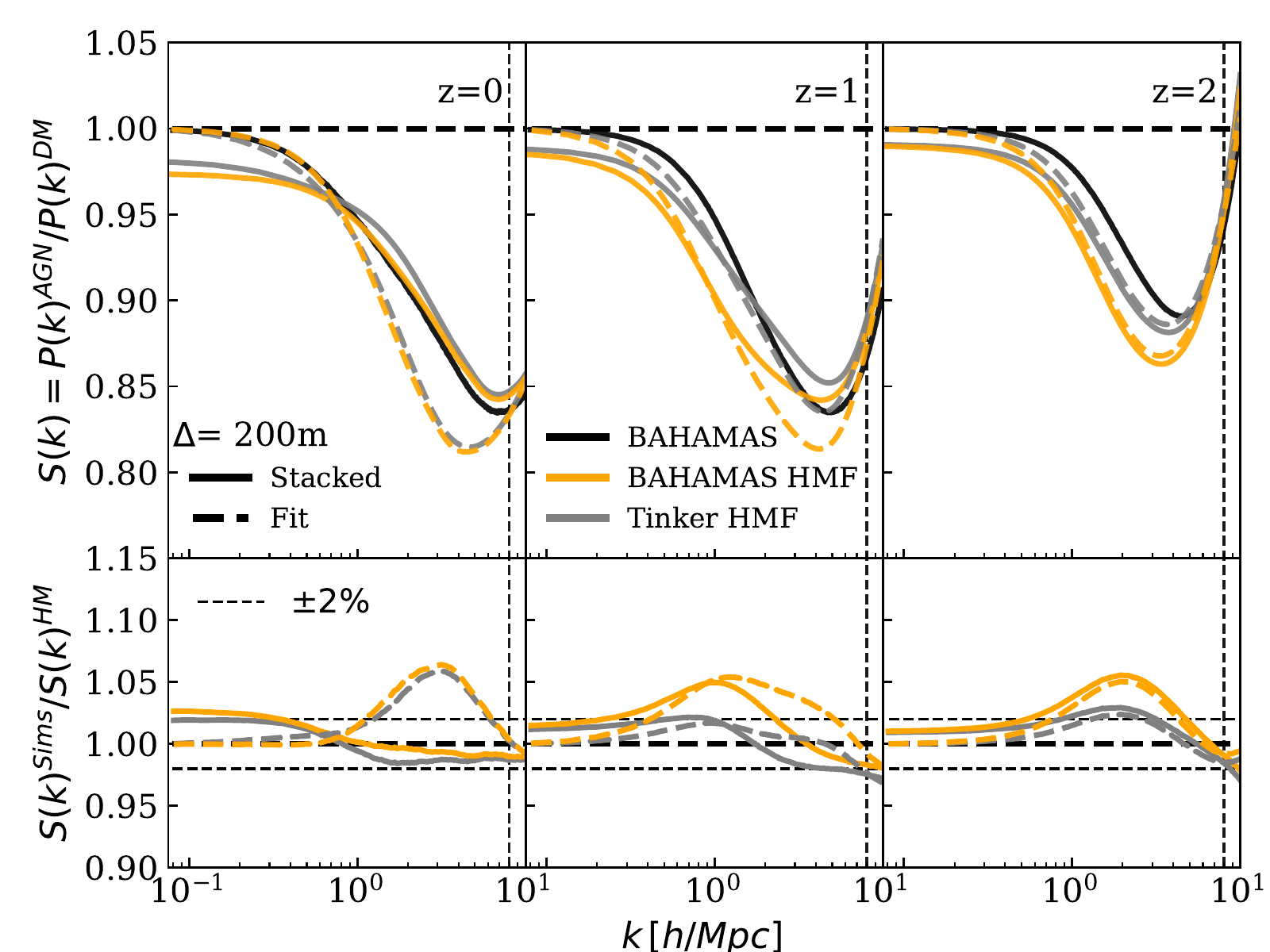}\par 
    \includegraphics[width=\linewidth]{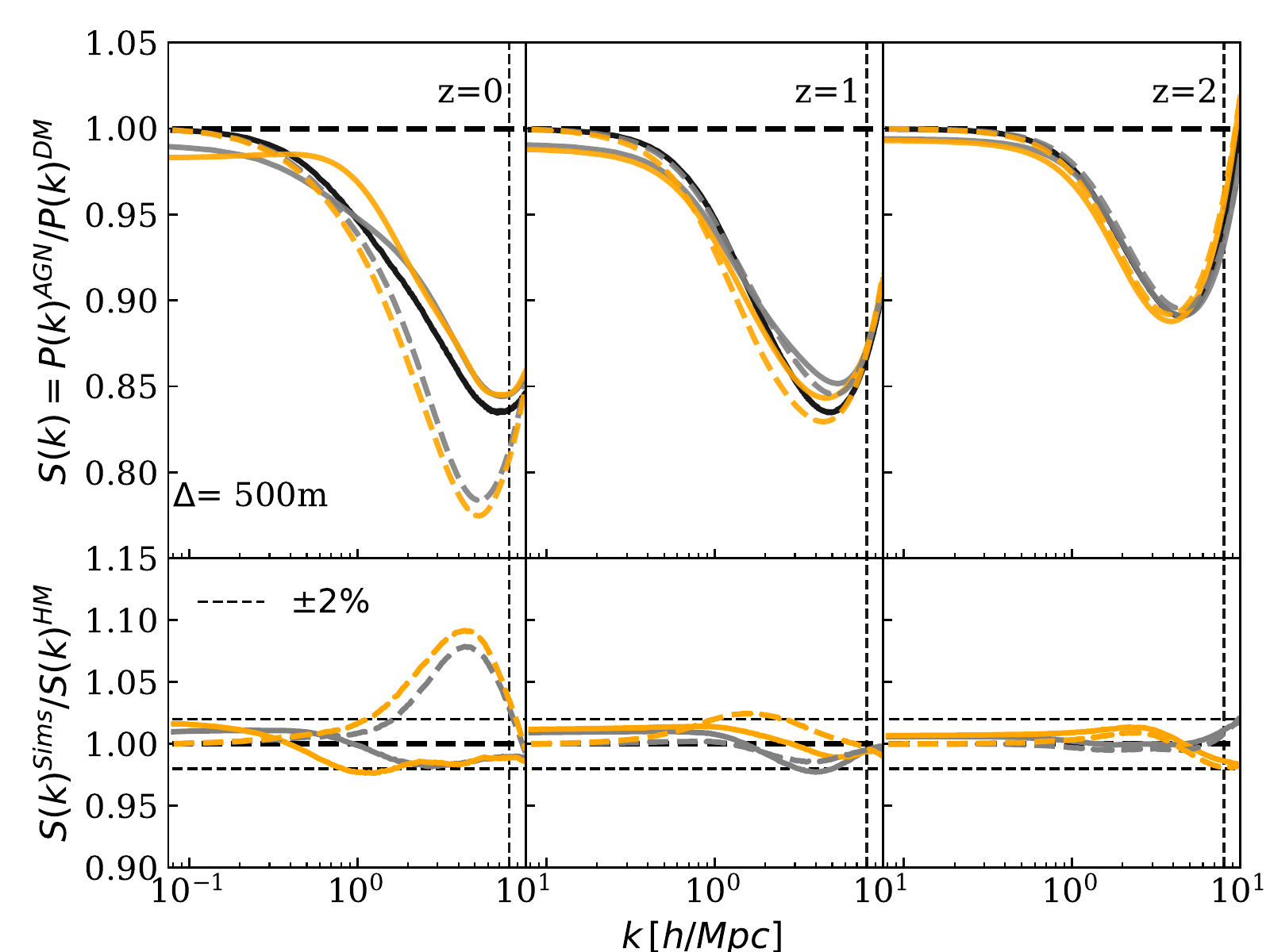}\par 
\end{multicols}
\begin{multicols}{2}
    \includegraphics[width=\linewidth]{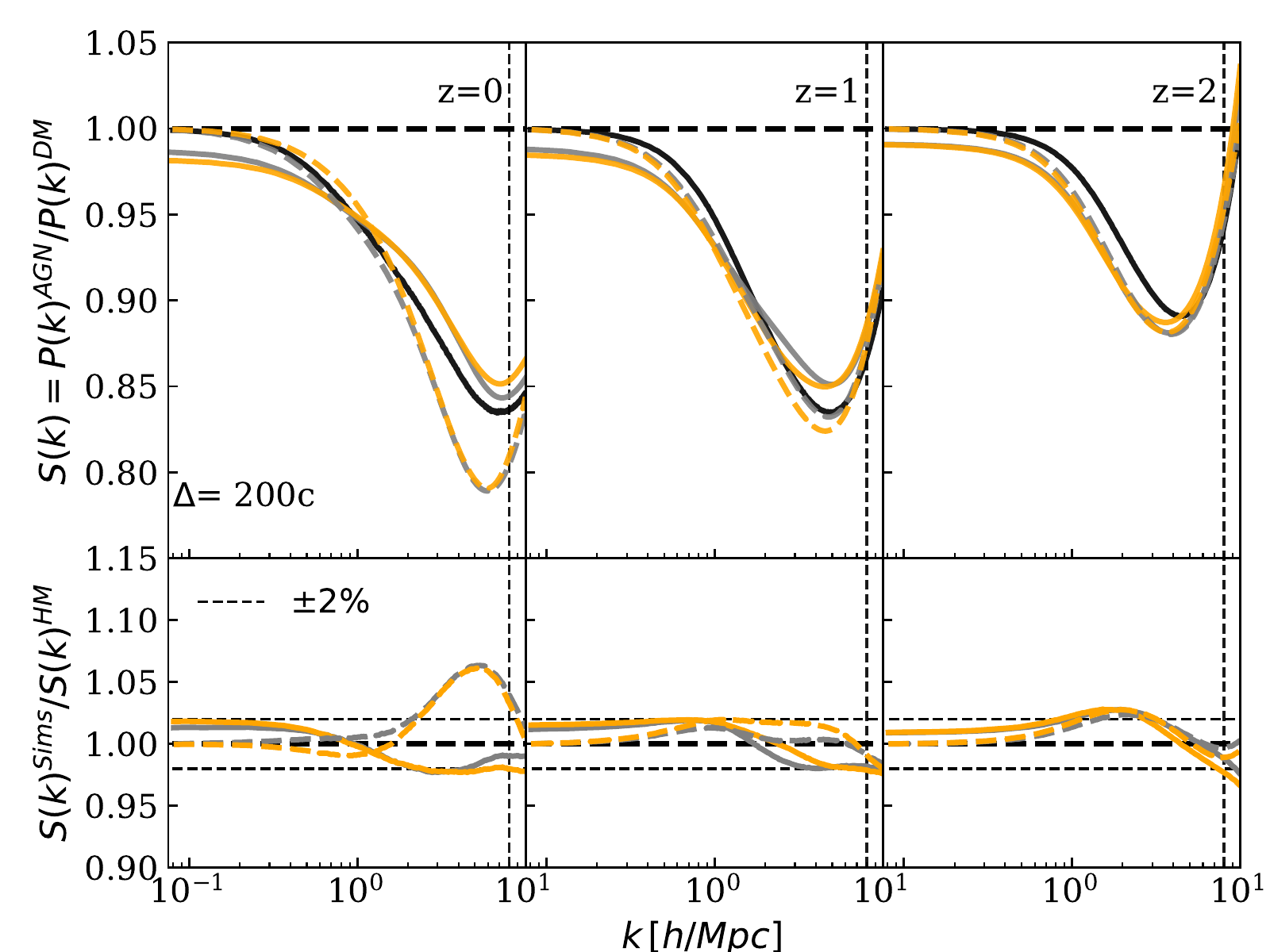}\par
    \includegraphics[width=\linewidth]{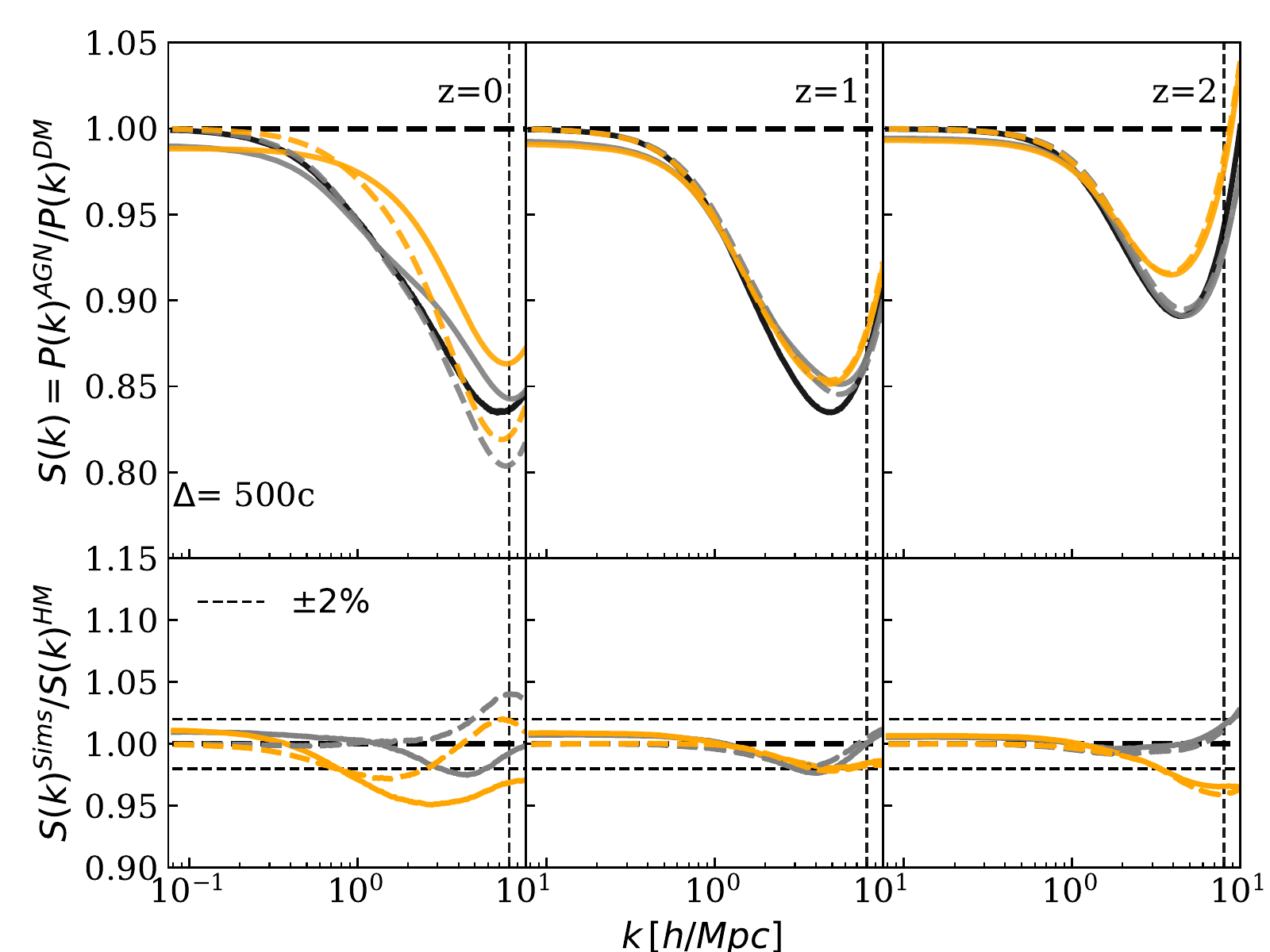}\par
\end{multicols}
\caption{Matter power spectrum suppression plots between the total matter power spectrum and the \dm~power spectra predictions using the halo model results at three different redshifts for the four different mass definition following the same structure as the \dm~and \agn~plots Fig.~\ref{fig:pk_ovr_stack_fit_results_dm} and Fig.~\ref{fig:pk_ovr_stack_fit_results_agn}. In the bottom panel we show the accuracy of our prediction with the ratio between the simulation expected results and the halo model prediction highlighting the $2\%$ difference using dashed black curves. The BAHAMAS predictions are shown in solid black curve. We show the prediction with BAHAMAS HMFs in solid orange curves while in grey we show the Tinker prediction, in dashed curves we show the case using the fit density profiles while in solid curves the stacked density profiles.  The halo model reproduces the simulated suppression factor to typically a few percent accuracy, independent of details such as the halo mass definition.}
\label{fig:pk_suppression}
\end{figure*}

In Fig.~\ref{fig:pk_suppression} we show the suppression effect of the baryons with respect to the \dm~simulations, where the suppression is defined simply as $S(k) \equiv P_{\textit{AGN}}(k)/P_{\textit{DM}}(k)$. The structure of the plots is similar to the previous ones that we have shown for the power spectra comparison (Figs.~\ref{fig:pk_ovr_stack_fit_results_dm} and \ref{fig:pk_ovr_stack_fit_results_agn}) but in this case we show the suppression power spectra (top panels) and the ratio between the BAHAMAS results and the halo model results (bottom panels). We show in solid (dashed) curves the predictions using the tabulated (fitted) density profiles. In orange we show the predictions using the BAHAMAS HMFs and in grey using the Tinker HMFs (with the baryonic correction applied to the \agn~cases).

On a qualitative level, we can see that the ratio of the baryon to collisionless halo models (top row of panels in each plot set) has a `spoon'-like form that closely mimics that found by taking the ratio of power spectra from hydrodynamical and collisionless simulations.  Examining the ratio of power spectrum suppression of the simulations with respect to that from the halo model (i.e., a ratio of ratios, in the bottom row of panels of each plot set in Fig.~\ref{fig:pk_suppression}), we can also see that there is no evidence of an issue of near the 1-halo/2-halo transition region, nor of any particular systematic issues as a function of halo mass definition or redshift.  Slight differences exist depending on which set of density profiles we use (tabulated vs. parametric), but it is nevertheless abundantly clear that the halo model formalism is considerably more accurate in predicting the matter power spectrum suppression factor, as opposed the absolute $P(k)$.  Typically, we find that the ratio of halo models is accurate at the $\approx$2-3 percent level.

\section{Summary and conclusions} \label{conclusions}

In this study we have assessed the accuracy of the halo model to predict the non-linear matter power spectrum, which is the basis of many large-scale structure cosmological probes.  The advantages of the halo model are its speed, flexibility, and its intuitive physical nature.  However, its accuracy in predicting the non-linear power spectrum needs to be carefully assessed and here we have posed a simple question: how well does the halo model predict the non-linear power spectrum, $P(k)$, from a cosmological simulation when the ingredients of the halo model (namely the halo mass function and mass density profiles) are extracted from the same simulation?  Although the question is simple, the test is in fact a demanding one, since once the mass function and density profiles (and cosmology) are specified, there are no free parameters in the standard halo model.  

We briefly summarise the main results below:
\begin{itemize}
\item We have computed the stacked (mean) total mass density profiles in bins of halo mass and redshift for the BAHAMAS \dm~and \agn~simulations (see Figs.~\ref{fig:stack_accuracy_dm} and \ref{fig:stack_accuracy_agn}, respectively) and provided Einasto profile fits to these profiles (see eqn.~\ref{Einasto_eq} and Table \ref{tab:fittings_prof}).
\item Using the BAHAMAS simulations, we have derived a correction to the halo mass function that encapsulates the presence and impact of baryons on haloes.  This correction works for every overdensity and up to $z=2$ with an accuracy better than $5\%$ (see Fig.~\ref{fig:HMF_BAR} and eqn.~\ref{velliscig_new}).  
\item Using density profiles and halo mass functions extracted from the BAHAMAS simulations, we have calculated the non-linear power spectrum, $P(k)$, using the standard halo model.  Qualitatively speaking, the standard halo model reproduces the power spectrum in both the collisionless and baryon cases (see Fig.~\ref{fig:comparison_result_DMonly} and Fig.~\ref{fig:comparison_result_AGN}, respectively) we have considered, correctly capturing both the large-scale, linear limit and the deep non-linear regime.
\item In detail, we find that the halo model struggles to quantitatively reproduce the simulation power spectrum on intermediate scales ($0.1 \la k [h/{\rm Mpc}] \la 5$) that mark the transition from the so-called 2-halo term (the clustering of nearby, correlated haloes) to the 1-halo term (the mass density distribution inside a single halo).  For example, at $z=0$ and adopting a halo mass defined with respect to 200 times the mean background density, the halo model predicts a $P(k)$ that is systematically lower than predicted by the cosmological simulations by up to 15-20\% (see Figs.~\ref{fig:pk_ovr_stack_fit_results_dm} and \ref{fig:pk_ovr_stack_fit_results_agn}). This result follows previous works (e.g. \citealt{massara14, voivodic2020}) but narrows down the source of uncertainties by using the halo mass function and density profiles directly from the simulations.
\item  We have shown that the choice of halo mass definition (defined with respect to the critical or mean background density and the choice of overdensity) has a significant impact on the 1-halo/2-halo transition region offset.  This effect is due to the change in the radial extent of the haloes depending on the mass definition, with larger radial extents (lower overdensities) generally resulting in an improved match between the halo model and the simulations.
\item The 1-halo dominated region is recovered to $5\%$ at $z=0$ and better than $10\%$ for all mass definitions, although the accuracy decreases at higher redshifts.  
\item While the standard (unmodified) halo model cannot predict the absolute power spectrum to better than 15\% accuracy on intermediate scales (at best), we have shown that these systematic errors largely cancel when considering the ratio of the baryon to collisionless cases.  Typically, the halo model can reproduce the suppression seen in the simulations to a few percent accuracy, independent of the details such as the halo mass definition (Fig.~\ref{fig:pk_suppression}).
\end{itemize}

One of the key findings of our study is that the accuracy of the halo model in reproducing the simulations is strongly affected by the halo mass definition, through its impact on the radial extent of haloes.  In essence, adopting higher overdensities implies smaller radial extents (for a given mass) which effectively confines the 1-halo contribution to smaller scales, resulting in lower power at the 1-halo/2-halo transition region and poorer agreement with the simulations.  One possibility is to simply radially extend the profiles associated with a given mass definition, as suggested recently by \citet{garcia2021}.  Alternatively, one can retain the link between the halo mass and radius and simply adopt a lower overdensity, or perhaps another physical scale (at typically low overdensities) such as the splashback radius.  In addition, \citet{meadverde} have shown that accounting for 
non-linear bias in the 2-halo term also helps to mitigate the error in the transition region.  Note that the standard halo model assumes a linear bias which is independent of scale, but in principle we expect the clustering to be scale-dependent on quasi-linear scales.

The other major finding of our study is that the ratio of power spectra (baryon case to collisionless case) can be much more robustly predicted with the standard halo model than can the absolute power spectra.  Interestingly, previous studies have similarly concluded that the effects of including massive neutrinos or of altering the nature of dark energy or gravity on the matter power spectrum are also most reliably captured with the halo model in terms of ratios (e.g., \citealt{schmidt2010,mead2017,cataneo2019,cataneo2020,bose21}).  In these studies, the ratio is sometimes referred to as the `response' or the `reaction' to a cosmological change.  Our results regarding the ratio of the baryon and collisionless halo models could therefore be termed as a `baryon response' or `baryon reaction'.  One possibility, is to use the halo model to predict the baryon response and combine this with other methods for computing the absolute power spectrum in the collisionless limit.  Fast and accurate emulators based on large suites of collisionless simulations are now readily available in an expanding cosmological parameter space (e.g., \citealt{lawrence2017,Euclid2021}).  Combining these with the flexible, physically-motivated halo model to account for the presence of baryons is a novel, interesting prospect and one that differs from existing methods to account for baryons, including those that use the halo model to compute the absolute power spectrum in the presence of baryons (HMcode; \citealt{Mead15,mead2020}), use analytic prescriptions for directly modifying the outputs of collisionless simulations such as the `baryonification' approach (\citealt{schneider2015,angulo2020, arico2020}), or that use full cosmological hydrodynamical simulations directly (BAHAMAS; \citealt{BAHAMAS}).

\section*{Acknowledgements}
The authors thank the referee for helpful suggestions that improved the paper.
The authors thank Alex Mead for helpful comments on the paper. The authors would like to acknowledge a LIV.DAT doctoral studentship supported by the STFC under contract[ST/P006752/1]. The LIV.DAT Centre for Doctoral Training (CDT) is hosted by the University of Liverpool and Liverpool John Moores University / Astrophysics Research Institute. SGS acknowledges an STFC doctoral studentship.  This project has received funding from the European Research Council (ERC) under the European Union’s Horizon 2020 research and innovation programme (grant agreement No 769130).    This work used the DiRAC@Durham facility managed by the Institute for Computational Cosmology on behalf of the STFC DiRAC HPC Facility. The equipment was funded by BEIS capital funding via STFC capital grants ST/P002293/1, ST/R002371/1 and ST/S002502/1, Durham University and STFC operations grant ST/R000832/1. DiRAC is part of the National e-Infrastructure.

\section*{Data Availability Statement}
The data underlying this article will be shared on reasonable request to the corresponding author.

\footnotesize{
\bibliographystyle{mnras}
\bibliography{halo_model}{}
}
 
\label{lastpage}

\end{document}